\documentclass[10pt,amsmath,aps,prb,superscriptaddress,amssymb,longbibliography,twocolumn]{revtex4-2}
\usepackage{xcolor}            
\usepackage{graphicx}
\usepackage[	
citecolor=blue,
colorlinks,
linkcolor=blue,
urlcolor=blue,
]{hyperref}
\usepackage{appendix}
\usepackage{wasysym}
\usepackage{slashed}
\usepackage{subfigure}
\usepackage{soul}
\usepackage{natbib}

\begin{document}
\title{From Quantum Dimers to the \texorpdfstring{$\pi$}{}-flux Toric Code via Deconfined Multicriticality}

\author{Ankush Chaubey}
\email{ankush.chaubey@icts.res.in}
 \affiliation{International Centre for Theoretical Sciences, Tata Institute of Fundamental Research, Bengaluru 560089, India.}
  \author{Sergej Moroz}
 \email{sergej.moroz@kau.se}
 \affiliation{Department of Engineering and Physics, Karlstad University, Karlstad, Sweden.}
 \affiliation{Nordita, KTH Royal Institute of Technology and Stockholm University, Stockholm, Sweden.}
  \author{Subhro Bhattacharjee}
 \email{subhro@icts.res.in}
 \affiliation{International Centre for Theoretical Sciences, Tata Institute of Fundamental Research, Bengaluru 560089, India.}
\date{\today}

\begin{abstract}
{Two-dimensional Rokhsar-Kivelson (RK) dimer models on bipartite lattices are generally limited to translation-symmetry-broken dimer crystals. We introduce a tensor-product regularisation of the dimer Hilbert space that yields a qubit Hamiltonian interpolating from the RK model to the $\pi$-flux toric code, thereby accessing a deconfined $\mathbb{Z}_2$ topological liquid. In this framework, the $\mathbb{Z}_2$ liquid descends from a multicritical $U(1)$ spin liquid through condensation of a charge-2 Higgs field, thus avoiding confinement. Using iDMRG together with low-energy field theory, we determine a phase diagram containing two continuous quantum phase transitions---a $3\mathrm{D}$ XY$^{\ast}$ transition between the $\mathbb{Z}_2$ liquid and the columnar/plaquette-VBS, and a quantum Lifshitz transition between two dimer crystals---alongside a first-order transition between the staggered crystal and the $\mathbb{Z}_2$ liquid. Our field theory suggests a deconfined multicritical point described by an Abelian Higgs model with dynamical critical exponent, $z=2$, where the three transitions meet, highlighting the interplay of fractionalisation and emergent gauge fluctuations.}
\end{abstract}

\maketitle

\section{Introduction}

Quantum dimer models (QDM) are interesting platforms for stabilizing unconventional correlated phases and phase transitions~\cite{anderson1973resonating,fazekas1974information,PhysRevB.35.8865,PhysRevLett.86.1881}. {In the context of spin systems,} dimer models can arise in frustrated spin-1/2 antiferromagnets where the spins locally minimize energy by forming spin singlets (dimers)~\cite{chayes1989valence,moessner2002resonating,PhysRevB.35.8865}, while the spinful excitations remain gapped. Such dimer models can also appear as an effective low-energy description of a variety of other systems, such as that of hard-core bosons~\cite{balents2002fractionalization,isakov2006hard,roychowdhury2015z} and, more recently, Rydberg atom arrays~\cite{samajdar2021quantum,yan2022triangular,weimer2010rydberg,PhysRevX.10.021057,PhysRevX.4.041037,PhysRevX.11.031005,PhysRevResearch.7.L012006}.

A paradigmatic Hamiltonian describing the low-energy physics of such dimers was introduced by Rokhsar and Kivelson in 1988~\cite{PhysRevLett.61.2376} on the square lattice in search of a short-ranged resonating valence bond (RVB)  liquid~\cite{anderson1987resonating,anderson1973resonating,anderson1974materials,PhysRevB.35.8865} in the context of cuprate superconductors. Disappointingly, however, the simple nearest neighbour RK model as well as related SU(N) spin models fail to realise such a liquid phase on the square and other two-dimensional bipartite lattices~\cite{PhysRevLett.61.2376,PhysRevB.40.5204,PhysRevLett.62.1694}, limiting their applicability to the physics of cuprate superconductivity envisaged as a doped RVB~\cite{anderson2004physics}. Indeed, subsequent extensive investigations of the  RK model and its various extensions on different lattices \cite{moessner2010quantum,PhysRevB.65.024504,PhysRevB.69.224416,PhysRevB.69.224415} in two dimensions show that they typically have an extended $\mathbb{Z}_2$ RVB liquid phase on non-bipartite lattices along with (spontaneously) lattice translation symmetry broken dimer crystal phases dubbed valence bond solids (VBS)~\cite{PhysRevLett.86.1881,PhysRevLett.89.137202,PhysRevB.68.054405}. However, on bipartite lattices, the liquid, if at all present, is fine-tuned to the special point in the parameter space -- dubbed the RK point -- whence the ground state is given by an equal superposition of all dimer coverings~\cite{henley2004classical,PhysRevLett.61.2376} while dimer crystals of different types extend over the rest of the phase diagram \cite{PhysRevLett.61.2376,PhysRevB.40.5204,PhysRevB.40.6941,PhysRevB.54.12938}.

In sharp contrast to the RK models, topologically ordered $\mathbb{Z}_2$ liquids are obtained on square and other bipartite (and non-bipartite) lattices in a different class of exactly solvable qubit Hamiltonians -- the toric code model \cite{kitaev2003fault}. The ground state and excitations of these $\mathbb{Z}_2$ liquids are effectively described as gapped Ising electric and magnetic charges of a deconfined phase of a IGT.

In this paper, we construct and analyze a simple microscopic spin-1/2 Hamiltonian on a square lattice that amalgamates the phenomenology of the QDMs -- the dimer crystals with critical liquid at the RK point, as well as allows for the extended $\mathbb{Z}_2$ liquid, as in the toric code model. Such a microscopic Hamiltonian provides rich settings for unconventional topological quantum phases as well as Landau forbidden quantum transitions~\cite{senthil2024deconfined,senthil2004deconfined} and hence an interesting playground for realising them in engineered platforms~\cite{samajdar2021quantum,yan2022triangular,weimer2010rydberg,PhysRevX.10.021057,PhysRevX.4.041037,PhysRevX.11.031005,PhysRevResearch.7.L012006}.

\begin{figure}
\includegraphics[width=0.9\linewidth]{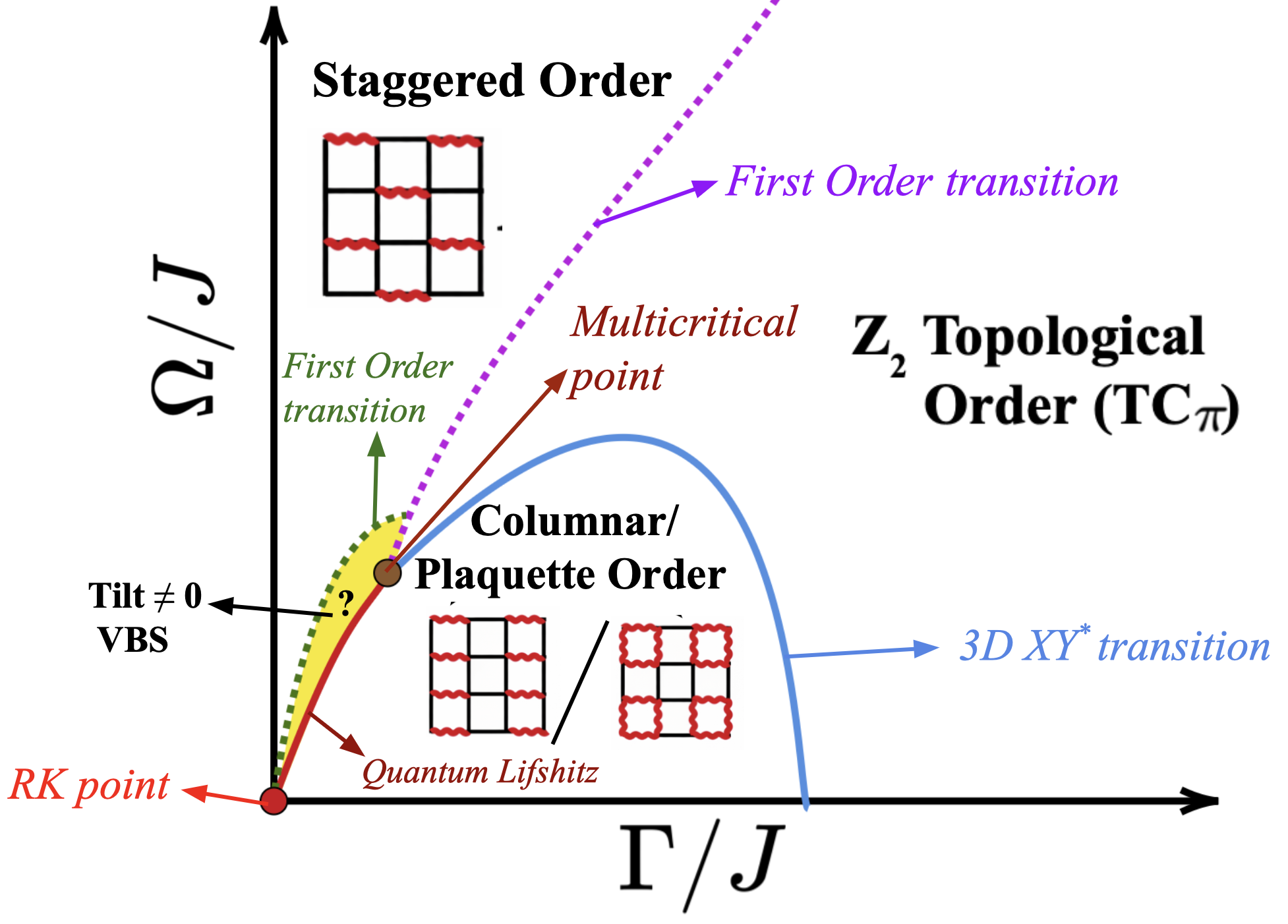}
\caption{{\bf Schematic phase diagram} of the model defined by the Hamiltonian in Eq. \eqref{eq_gauge_invariant_H_deconf_full} exhibiting three major phases: $\mathbb{Z}_2$ topologically ordered (TC$_\pi$), c/p-VBS and s-VBS, separated by three phase boundaries -- (i) 3D $XY^{*}$ critical line between the c/p-VBS and the $\mathbb{Z}_2$ liquid, (ii) a quantum Lifshitz transition between the two VBSs, and (iii)  first-order transition between the s-VBS and $\mathbb{Z}_2$ liquid. All these phase transitions meet at a multicritical point. The yellow shaded region corresponds to a small region of possible incommensurate VBS phase with a finite tilt expected from the field theory. {The question mark (``?") for the tilted VBS indicates that the present numerics do not provide conclusive evidence for this phase.} }
\label{fig_phase_diagram}
\end{figure}

Key to our construction is a particular qubit regularisation of the QDM to obtain a generalised spin-1/2 Hamiltonian (Eq. \eqref{eq_gauge_invariant_H_deconf_full}) that reduces to the QDM and the toric code in appropriate limits. A combination of analytical and numerical results lead to a rich phase diagram that is illustrated in Fig. \ref{fig_phase_diagram}, consisting of {the $\mathbb{Z}_2$ topological liquid}  and two confined symmetry broken phases -- the staggered VBS {(s-VBS)} and the columnar/plaquette VBS {(c/p-VBS) (Fig. \ref{fig_columnar_large_omega})}. {Notably, on a torus, the c/p-VBS and the s-VBS belong to different winding number sectors -- while the former lies in a zero tilt (winding number per unit length) sector, the latter belongs to the maximally tilted sector~\cite{moessner2010quantum,fradkin2013field}}.

Our Density-Matrix renormalisation Group (DMRG) calculations on an infinitely long cylinder with finite perimeter efficiently capture the two dimer crystals via their respective order parameters and the $\mathbb{Z}_2$ liquid via the topological entanglement entropy. Combining this with the correlation length data, we approximately locate the position of the phases and various phase transitions. 

A convenient starting point to understand the phase transitions -- several of them Landau-forbidden --  are the two anyonic excitations of the $\mathbb{Z}_2$ liquid stabilised in the toric code limit of our microscopic Hamiltonian (Eq.~\eqref{eq_gauge_invariant_H_deconf_full})- the bosonic $\mathbb{Z}_2$ electric and magnetic charges with mutual semionic statistics~\cite{kitaev2003fault}. While these charges are gapped in the topological liquid, the electric charges condense in some parameter regime, resulting in VBS order. The exact nature of the VBS order depends on the momentum of the soft electric modes. The basic building block of the various critical theories is obtained from the transition between the $\mathbb{Z}_2$ liquid and the c/p-VBS that is brought about by the condensation of a pair of soft electric modes~\cite{PhysRevB.30.1362,PhysRevB.44.686,PhysRevB.84.104430} at commensurate momenta. These soft modes transform (projectively) under various microscopic lattice symmetries to lead to an enhanced $O(2)$ symmetry at low energy continuum limit at the critical point such that the transition between the $\mathbb{Z}_2$ liquid and the c/p-VBS is described by an Abelian Higgs model with a mutual $U(1)$ Chern-Simons term and belongs to a 3D XY$^*$ universality class~\cite{PhysRevB.44.686,PhysRevB.30.1362}.  

The obtained Abelian Higgs model, however, has a much more general applicability, and, when supplemented with the right projective symmetry, allowed higher order gradient terms (Eq.~\eqref{eq_lifshitz}) -- the so-called Lifshitz terms -- can describe the transition between various VBS crystal phases. In fact, in the regime dominated by the phase fluctuation of the soft electric mode, our field theory reduces to the critical theories derived starting with the height model representations of the QDM~\cite{fradkin2013field,PhysRevB.69.224415,PhysRevB.69.224416,ardonne2004topological}. As a consequence, in this regime, our results agree with earlier works~\cite{PhysRevB.69.224416,PhysRevB.69.224415}-- the transition between the two dimer crystals can generically occur either via a direct first-order transition or multiple steps with VBS phases having intermediate tilt via continuous quantum Lifshitz transition and the first-order line with the quantum Lifshitz lines being separated by an RK point, see Fig. \ref{fig_fpd}. We expect that for our specific microscopic model, we should see the multi-step transition as indicated in the Fig. \ref{fig_phase_diagram} and argued below. {However, in our DMRG calculations—limited by the finite circumference of the cylinder—we do not observe any definitive signatures of a partially tilted phase or an incomplete {\it devil's staircase}, as argued in Refs. \onlinecite{PhysRevB.69.224416,PhysRevB.69.224415}. Nevertheless, a narrow parameter window between the c/p-VBS and s-VBS phases suggests the presence of a partially tilted VBS phase, highlighted by the yellow shaded region in Fig. \ref{fig_phase_diagram}.}

Further, the same Lifshitz terms, in an appropriate regime, lead to the deviation of the momenta of the soft electric modes away from the c/p-VBS and lead to a direct first-order transition between the $\mathbb{Z}_2$ liquid and VBSs with finite tilt, including the maximally tilted s-VBS, as shown in Fig.~\ref{fig_phase_diagram}.

The continuum Abelian Higgs theory, consistent with symmetries of the microscopic model, allows for a deconfined multicritical point, obtained by tuning two coupling constants, where the three transition lines meet; see Figs. \ref{fig_phase_diagram} and \ref{fig_fpd}-- (i) the 3D XY$^*$ line between the $\mathbb{Z}_2$ liquid and the c/p-VBS, (ii) the continuous $(2+1)D$ quantum Lifshitz transition between the c/p-VBS (with zero tilt) and VBS with finite tilt, and, (iii) the first-order transition between the $\mathbb{Z}_2$ liquid and the s-VBS. At this point, a gapless multicritical $U(1)$ liquid is realised with dynamical critical exponent, $z=2$, somewhat similar to the RK point, but distinct from it. From the perspective of the multicritical $U(1)$ liquid, the $\mathbb{Z}_2$ liquid is naturally obtained, as we show, by condensation of the charge-two Higgs scalar~\cite{sachdev1999translational} which is dual to the electric charges of the toric code. On the other hand, the two dimer crystals correspond to the two different confined phases of the critical $U(1)$ liquid~\cite{fradkin2013field,polyakov81003gauge,ardonne2004topological}. The resultant physics fall beyond the purview of the standard Landau-Ginzburg-Wilson paradigm of critical phenomena and provide a concrete mechanism to stabilise a $\mathbb{Z}_2$ liquid in the vicinity of the dimer manifold on a bipartite lattice.

The rest of the paper is organised as follows. In Sec. \ref{sec_ham}, starting with the RK model, we write down a spin-1/2 Hamiltonian which interpolates between the RK quantum dimer limit and the toric code with a background $\pi$-flux. We also discuss analytically tractable corners of the phase diagram. In Sec. \ref{sec_numerics}, we report results of our numerical DMRG simulations performed on infinite cylinders, which provide a quantitative confirmation of the phase diagram shown in Fig. \ref{fig_phase_diagram}. In Sec. \ref{sec_EFT}, we develop a low-energy field theory that captures all quantum phases and associated phase transitions. We draw our conclusions and provide an outlook in Sec. \ref{sec_summary}. Various technical details are summarised in the appendices.

\section{The model and its limits}
\label{sec_ham}

\begin{figure}
\includegraphics[scale=0.45]{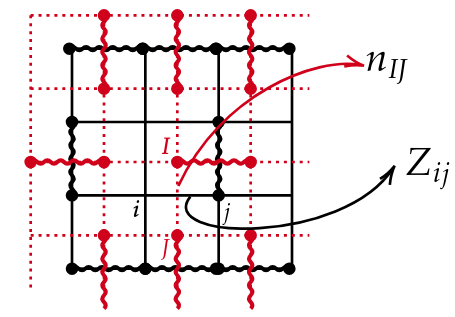}
\caption{{\bf Mapping between dimers and Ising variables (Eq. \eqref{eq_dimer_mapping}) :}
The red bonds with curly (dotted) lines represent the presence (absence) of dimers (Eq. \eqref{eq_gaussdimer}) residing on a square lattice.
The black curly (solid) bonds  represent  Ising variable $Z=-1(Z=+1)$ on bonds of the dual lattice. }
    \label{fig_dimer_model}
\end{figure}

Central to the richness of the QDMs is the fact that the dimer Hilbert space does not have a tensor product structure. This arises from the fact that each spin can be a part of one dimer. Indeed, on a square lattice with $N$ sites, the number of hard-core dimer coverings scales as $\sim (1.339)^N$~\cite{fisher1961statistical,kasteleyn1961statistics}. The dimer models are described in terms of dimer numbers, $n_{IJ}=0,1$, on each bond connecting the sites $I$ and $J$ on a lattice (see Fig. \ref{fig_dimer_model}), such that the single dimer attached to each site $I$ is given by the constraint
\begin{align}
    \sum_{J \in I} n_{IJ} = 1~~~~~~~~\forall~I,
    \label{eq_gaussdimer}
\end{align}
where the sum is over all the sites, $J$, connected to $I$ on the lattice. On a square lattice, the typical local dimer dynamics, maintaining the above constraint and the topological sectors~\cite{moessner2010quantum}, is captured by the paradigmatic RK Hamiltonian~\cite{PhysRevLett.61.2376}
\begin{align}
    H_{\text{Dimer}} & =-\Gamma\sum\left(|\vcenter{\hbox{\includegraphics[height=0.02\textheight]{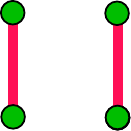}}}\rangle\langle\vcenter{\hbox{\includegraphics[height=0.02\textheight]{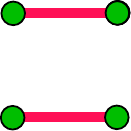}}}|+|\vcenter{\hbox{\includegraphics[height=0.02\textheight]{QDMTerms/new_term_2.pdf}}}\rangle\langle\vcenter{\hbox{\includegraphics[height=0.02\textheight]{QDMTerms/new_term_1.pdf}}}|\right) \nonumber \\
    &+ \Omega\sum\left(|\vcenter{\hbox{\includegraphics[height=0.02\textheight]{QDMTerms/new_term_1.pdf}}}\rangle\langle\vcenter{\hbox{\includegraphics[height=0.02\textheight]{QDMTerms/new_term_1.pdf}}}|+|\vcenter{\hbox{\includegraphics[height=0.02\textheight]{QDMTerms/new_term_2.pdf}}}\rangle\langle\vcenter{\hbox{\includegraphics[height=0.02\textheight]{QDMTerms/new_term_2.pdf}}}|\right),
    \label{eq_RK_hamiltonian}
\end{align}
where the first (second) term provides kinetic (potential) energy to the dimers. 

To generalize the above dimer model on the square lattice (drawn in red dotted lines in Fig.~\ref{fig_dimer_model}), we associate with each bond where a dimer resides, an Ising spin $Z_{ij}=\pm 1$, where $ij$ refers to the uniquely associated bond of the dual square lattice (shown in black), such that
\begin{align}\label{eq_dimer_mapping}
    n_{IJ}=\frac{1-Z_{ij}}{2}.
\end{align}
The single-dimer constraint Eq.~\eqref{eq_gaussdimer}, expressed in terms of the Ising spins, becomes
\begin{align}
    \sum_{\langle ij \rangle \in \square} Z_{ij} = 2,
    \label{eq_gaussising1}
\end{align}
where the sum is over each elementary dual square plaquette enclosing one site of the direct lattice.

The QDM Hamiltonian~\eqref{eq_RK_hamiltonian}, written in terms of the Ising spins, then takes the form 
\begin{align}
    \mathcal{H}_{\rm Dimer} =& -\Gamma \sum_i \prod_{j\in i} X_{ij}\nonumber\\
    &+\frac{\Omega}{4}\sum_{i}\left[(1-Z_{i-\hat{x},i})(1-Z_{i+\hat{x},i})\right.\nonumber\\
    &\qquad\qquad\qquad\left.+(1-Z_{i-\hat{y},i})(1-Z_{i+\hat{y},i})\right],
\end{align}
where $X_{ij}$ anticommutes with $Z_{ij}$ and induces spin flips. The constraint in Eq.~\eqref{eq_gaussising1} can now be implemented by adding an energy-cost term to the Hamiltonian of the form
\begin{align}
    \mathcal{H}_{\rm constraint} = \kappa\sum_{\square}\prod_{\langle ij\rangle\in\square} Z_{ij}
    - J\sum_{\langle ij\rangle} Z_{ij},
    \label{eq_gauge_invariant_H_deconf}
\end{align}
where both $\kappa, J>0$ with $\kappa\gg J$. The first term, in the limit $\kappa\rightarrow\infty$, enforces $\pi$-flux
\begin{align}
    \prod_{\langle ij \rangle \in \square} Z_{ij} = -1,
    \label{eq_gaussising2}
\end{align}
on each dual-lattice plaquette. From Eq.~\eqref{eq_dimer_mapping}, this corresponds to having one or three dimers per site of the direct lattice. This degeneracy is lifted by the second term in Eq. \eqref{eq_gauge_invariant_H_deconf}, leaving only one dimer per site. {Throughout this work, we set $\kappa, \Gamma > 0$.}

Putting everything together results in the spin-$1/2$ Hamiltonian
\begin{align}
\mathcal{H} &= \mathcal{H}_{\rm constraint} + \mathcal{H}_{\rm Dimer}.
    \label{eq_gauge_invariant_H_deconf_full}
\end{align}

This resulting qubit Hamiltonian has a tensor product Hilbert space and incorporates the dimer regime ($\kappa\gg J\gg \Gamma, \Omega$) discussed above, with the point $\Gamma=\Omega=0$ corresponding to the classical dimer model with a macroscopically degenerate ground-state manifold. Also, a finite $J$ allows for the mixing between the one and the three-dimer states.

\begin{figure}
\includegraphics[width = 1.00\linewidth]{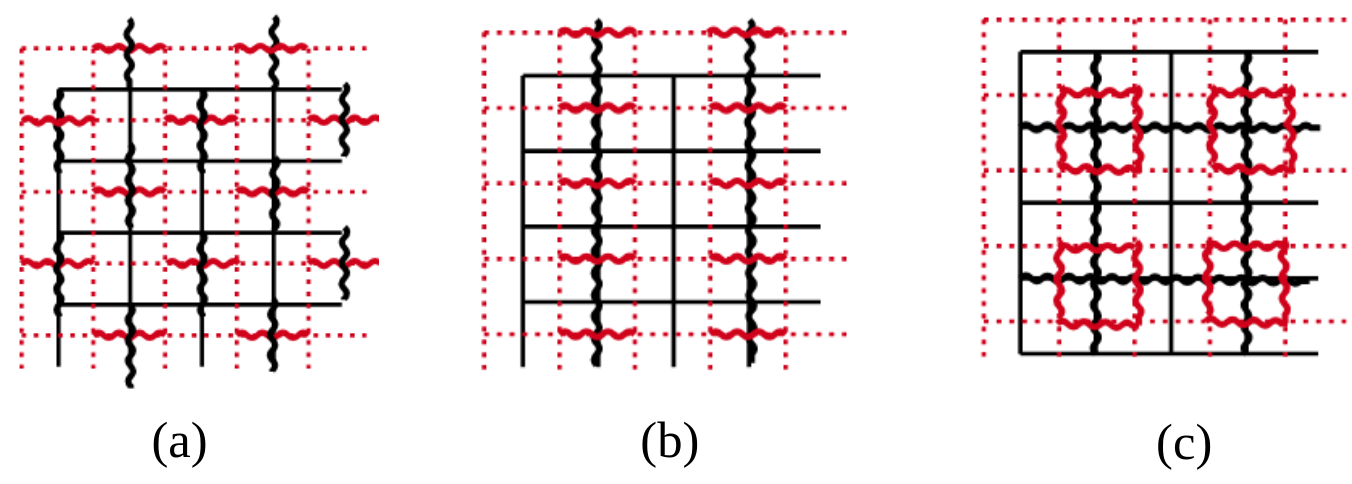}
\caption{{\bf Different dimer crystals:} (a) staggered, (b) (vertical) columnar, and  (c) plaquette VBSs.
}
    \label{fig_columnar_large_omega}
\end{figure}

Along the $\Gamma=0$ and $\Omega>0$ line, the model remains classical, but, the extensively degenerate classical dimer manifold splits energetically favouring the staggered configurations (Fig. \ref{fig_columnar_large_omega}(a)) locally characterised by $\prod_{\langle ij \rangle \in \square}Z_{ij} = -1$ and  $(1-Z_{i,i-{\bf n}})(1-Z_{i,i+{\bf n}})=0$, where ${\bf n}\in \{{ \bf \hat x},{\bf \hat y}\}$. Notably, in this classical limit, the domain walls of such staggered arrangements are free of energy cost, leading to a sub-extensively degenerate staggered manifold characterised by local s-VBS order (Fig.~\ref{fig_columnar_large_omega}(a)) with domain walls. 

{Quantum fluctuation ($\Gamma\neq 0$) lead to {\it order-by disorder} resulting in dimer crystals. This is expected as QDMs of bipartite lattices are described by pure compact $U(1)$ gauge theories~\cite{sachdev1999translational,PhysRevB.42.4568}, which, in $(2+1)$ dimensions are always confined~\cite{polyakov81003gauge}. Thus, in the dimer regime ($\Gamma/J,\Omega/J\ll 1$), we get a s-VBS phase (Fig.~\ref{fig_columnar_large_omega}a) when the potential term dominates ( $\Omega>\Gamma$), and c/p-VBSs (Figs.~\ref{fig_columnar_large_omega}b and \ref{fig_columnar_large_omega}c)  when kinetic term dominates ($\Gamma >\Omega $).}

{The two crystal phases are separated by a phase transition corresponding to the RK point as indicated in Fig. \ref{fig_phase_diagram}. Indeed, on taking $(\Gamma/J,\Omega/J)\rightarrow (0,0)$ at finite $\Gamma$ and $\Omega$ via $\kappa, J\rightarrow\infty$~\footnote{Note that this is different from approaching the origin by by taking $\Gamma,\Omega\rightarrow 0$ at finite $J$ whence we get a classical dimer model as mentioned above. Therefore the origin of the phase diagram in Fig. \ref{fig_phase_diagram} corresponding to $(\Gamma/J, \Omega/J)=(0,0)$ is singular.}, we recover the QDM described by the RK Hamiltonian (Eq. \eqref{eq_RK_hamiltonian}) to the leading order in perturbation theory in $\Gamma/J$ and $\Omega/J$ from Eq. \eqref{eq_gauge_invariant_H_deconf_full}. Thus, on approaching the origin along the $\Gamma=\Omega$ line, we reach an RK point that corresponds to the infinitely fine-tuned gapless $U(1)$ liquid (with quadratically dispersing photons) that describes a deconfined quantum phase transition between the c/p-VBS and the s-VBS~\cite{fradkin2013field,henley2004classical,PhysRevB.65.024504,ardonne2004topological}.}

In principle, as pointed out in Refs. \onlinecite{PhysRevB.69.224416,PhysRevB.69.224415}, the transition between the zero tilt c/p-VBS and the maximally {\it tilted} s-VBS need not be direct and can be intervened by a series of intermediate tilt VBS phases via an incomplete devil's staircase. {Our numerics (see below) indicate a very small sliver where both the s-VBS and c/p-VBS order parameters are zero, but limitations of finite size and aspect ratio cannot reveal the nature of this phase and its fate in the thermodynamic limit-- in particular, if it corresponds to a series of VBS phases with intermediate tilt.} {This is also expected from the perturbation theory in $\Gamma/J$ and $\Omega/J$ around the RK point, which leads to subleading corrections to the RK Hamiltonian and hence destroying the fine-tuning needed to achieve the ground state RK wave-function.}

On the other hand, for $\kappa, \Gamma\gg J, \Omega$, the Hamiltonian (Eq. \eqref{eq_gauge_invariant_H_deconf_full}) reduces to the toric code model~\cite{kitaev2003fault}
\begin{align}
    \mathcal{H}_{\rm TC}=&\kappa\sum_{\square}\prod_{\langle ij\rangle\in\square}Z_{ij}-\Gamma\sum_i \left[\prod_{j\in i}X_{ij}\right],
    \label{eq_tc}
\end{align}
that realises the $\mathbb{Z}_2$ topological liquid as an exact ground state since the two sets of terms (stabilisers) in the above Hamiltonian commute. The two kinds of gapped bosonic excitations -- magnetic and electric -- obtained by violating the first and second stabilisers, respectively, see each other as sources of the $\pi$-flux. The resulting mutual anyonic statistics manifests $\mathbb{Z}_2$ quantum topological order with four-fold degenerate ground state on a torus.

Hence, in the toric code limit in the ground state, one finds
\begin{align} \label{eq:pi-flux}
    \prod_{\langle ij\rangle\in\square} Z_{ij} = -1 \qquad \forall\, \square
\end{align}
and
\begin{align}
    \prod_{j\in i} X_{ij} = +1 \qquad \forall\, i,
\end{align}
We refer to this as the $\pi$-flux toric code (TC$_\pi$), since each electric charge experiences a $\pi$-flux when going around a closed contour containing an odd number of plaquettes. {This leads to momentum fractionalisation for the electric charge, but not for the magnetic charge.} This should be distinguished from the ``odd'' toric code studied recently in Ref. \onlinecite{borla2025odd}, where both plaquettes and stars have negative vacuum expectation values.

In the rest of this paper, we focus on the $\kappa \rightarrow \infty$ limit of the above Hamiltonian, which imposes the hard $\pi$-flux constraint \eqref{eq:pi-flux}, resulting in infinite energy cost of magnetic charges while the electric charges have energy of order $\Gamma$. The quantum phase diagram becomes a function of two dimensionless parameters that we choose to be $\Gamma/J$ and $\Omega/J$ in Fig.~\ref{fig_phase_diagram}.

Moving away from the TC limit along the $\Omega=0$ line, the finite $J$ term acts as a Zeeman field that makes the electric charges dynamic, resulting in a dispersion with bandwidth of order $J$. For $J\sim \Gamma$, the electric charges condense. Due to the background $\pi$-flux seen by the electric charges, upon condensation \footnote{At this point, we comment on the convention used in the present work relative to the existing dimer-model literature. In earlier works~\cite{PhysRevB.62.7850,sachdev1990effective,fradkin2013field,PhysRevB.84.104430}, crystalline phases are usually understood as vison condensates, where the vison corresponds to the Ising magnetic flux associated with the plaquettes of the lattice on whose bonds the dimers reside (see Fig.~\ref{fig_dimer_model}). However, these plaquettes correspond to sites of the dual lattice, on which—following the convention of the toric code Hamiltonian~\cite{kitaev2003fault}—the excitations are called electric charges. It is important to keep this nomenclature in mind to avoid confusion.} two finite-momentum soft modes develop, leading to a translation-symmetry-broken phase, namely the c/p-VBS. The critical theory describes a Landau-forbidden continuous phase transition between a topologically ordered $\mathbb{Z}_2$ liquid phase and a translation-symmetry-broken c/p-VBS. After duality transformation (Appendix \ref{appen_z2gt}), the lattice model reduces to the frustrated Ising model whose phase transition belongs to the 3D XY$^*$ universality class~\cite{PhysRevB.30.1362} as discussed below.

The gapped $\mathbb{Z}_2$ topological phase is clearly stable to turning on small $\Omega$. However, deep inside the $\mathbb{Z}_2$ liquid, increasing $\Omega$ eventually leads to the potential-energy–dominated s-VBS phase through a first-order phase transition (see below). The remarkable stability of the s-VBS at large $\Omega/J$ for all values of $\Gamma/J$, as well as its eventual first-order transition to the topological liquid, can be traced to the absence of $\Gamma$ perturbation-induced local dynamics within the s-VBS phase.

Finally, we notice that the constraint \eqref{eq:pi-flux} corresponds to the odd Ising Gauss law on the original lattice, shown in red in Fig. \ref{fig_dimer_model}. Owing to a quantum anomaly associated with translational symmetry, the odd Ising gauge theory (IGT) on a square lattice cannot support a trivially gapped phase with a unique ground state \cite{sachdev2018topological}. This obstruction is clearly reflected in the phase diagram shown in Fig. \ref{fig_phase_diagram}.

Having fleshed out the limits and the general structure of the phase diagram, we now turn to numerical simulations to obtain a quantitative confirmation of the phases as well as the associated phase transition lines. 

\section{Numerical Results}
\label{sec_numerics}

We performed infinite DMRG (or iDMRG) calculations in the infinite cylinder geometry using the tensor network Python (TeNPy) library~\cite{SciPostPhysLectNotes.5}. In this approach, the two-dimensional lattice sites are mapped onto a one-dimensional infinite chain by using a ``snake" path that traverses the cylinder circumference. This mapping allows us to exploit the efficiency of DMRG in one dimension, while still capturing the essential two-dimensional physics on a cylinder of finite circumference. The iDMRG algorithm~\cite{mcculloch2008infinite} works directly in the thermodynamic limit. Rather than simulating a finite system, the method optimises the tensors of an infinite Matrix Product State (iMPS) with a repeating unit cell. A key advantage of iDMRG is the ability to extract physical quantities from the MPS representation of the ground state. One of them is the MPS correlation length $\xi$, which can be obtained from the eigenvalue spectrum of the MPS transfer matrix~\footnote{The MPS is normalised such that the absolute value of the largest eigenvalue of the transfer matrix is unity, $|\lambda_1|=1$. In this normalisation, the MPS correlation length $\xi$ is determined by the second-largest eigenvalue $\lambda_2$ ~\cite{SciPostPhysLectNotes.5} via the formula $\xi = -1/\ln(|\lambda_2|)$.}. The MPS correlation length determines the decay of the slowest decaying correlation function \cite{SciPostPhysLectNotes.5}. Since $\xi$ diverges in gapless systems, we can identify quantum critical points by observing peaks in $\xi$ that grow with the MPS bond dimension, $\chi$. This provides a clear signature of criticality without requiring further system details. In our numerical simulations, we work on an infinite cylinder of circumference equal to four links, $L_y = 4$. We set $\kappa = 10$ as the largest coupling constant. In addition, we fixed $J = 1$ and varied the parameters $\Gamma$ and $\Omega$.

\begin{figure}
\begin{subfigure}[]{
\includegraphics[scale=0.11]{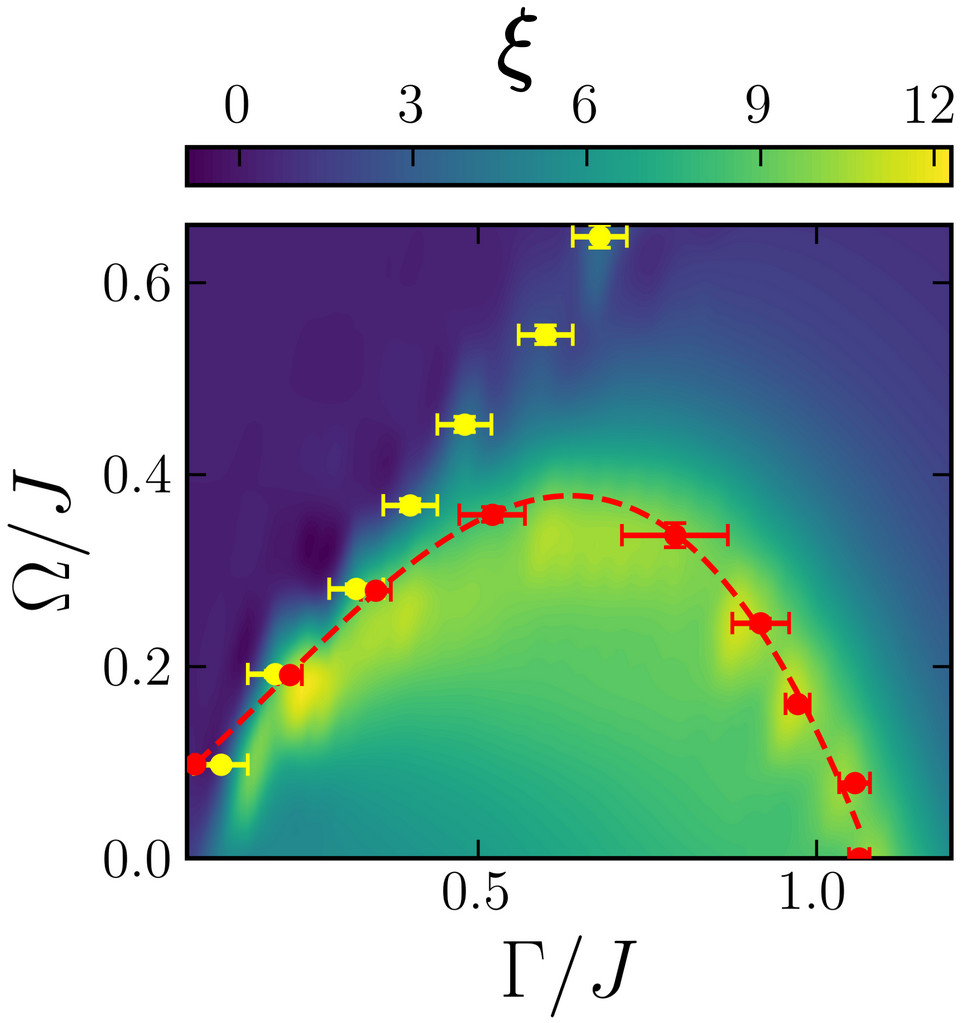}
}
\end{subfigure}
\begin{subfigure}[]{
\includegraphics[scale=0.11]{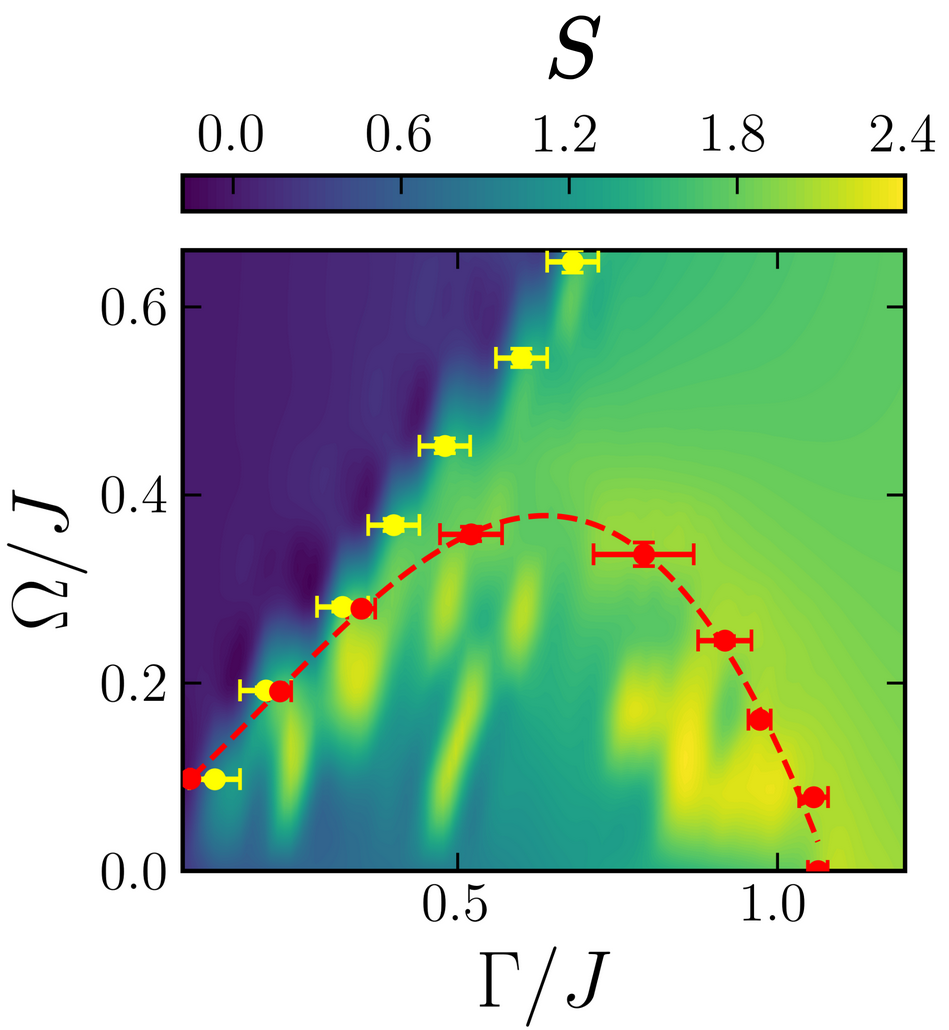}
}
\end{subfigure}
\begin{subfigure}[]{
    \includegraphics[scale=0.11]{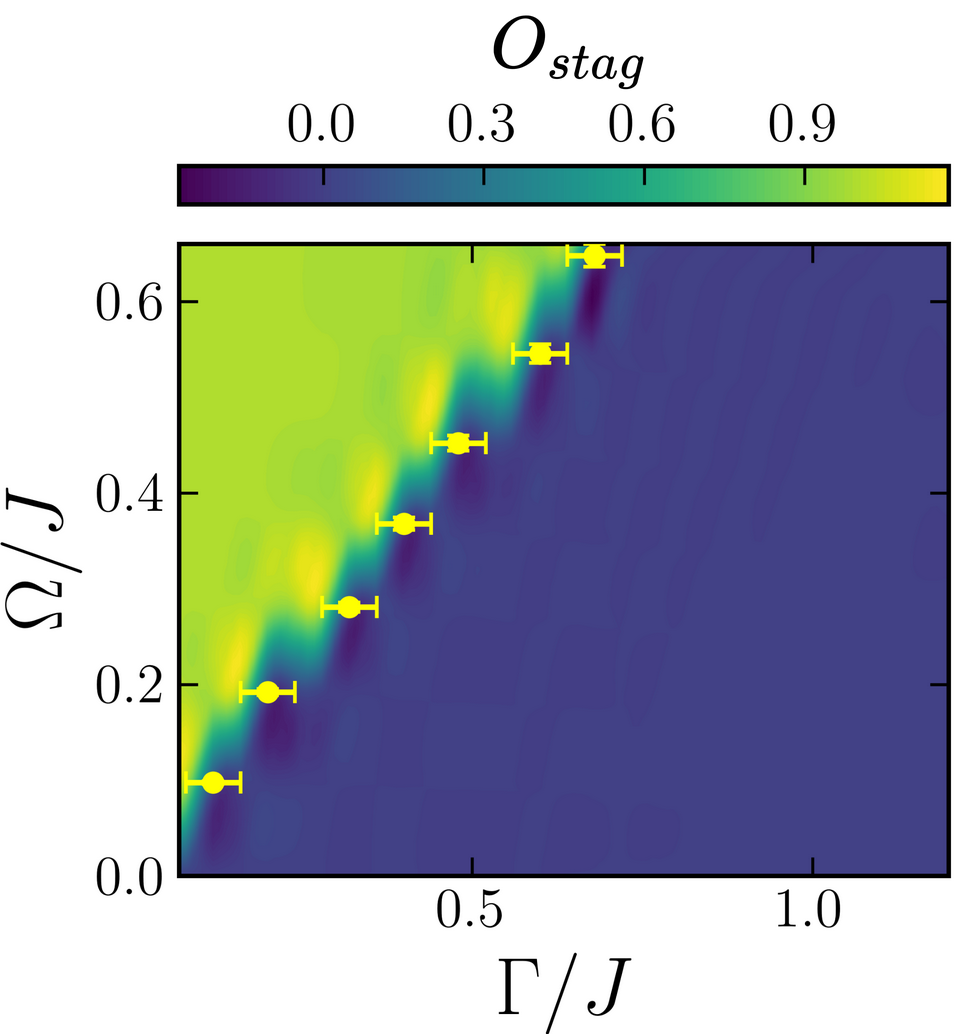}
    }
\end{subfigure}
\begin{subfigure}[]{
\includegraphics[scale=0.11]{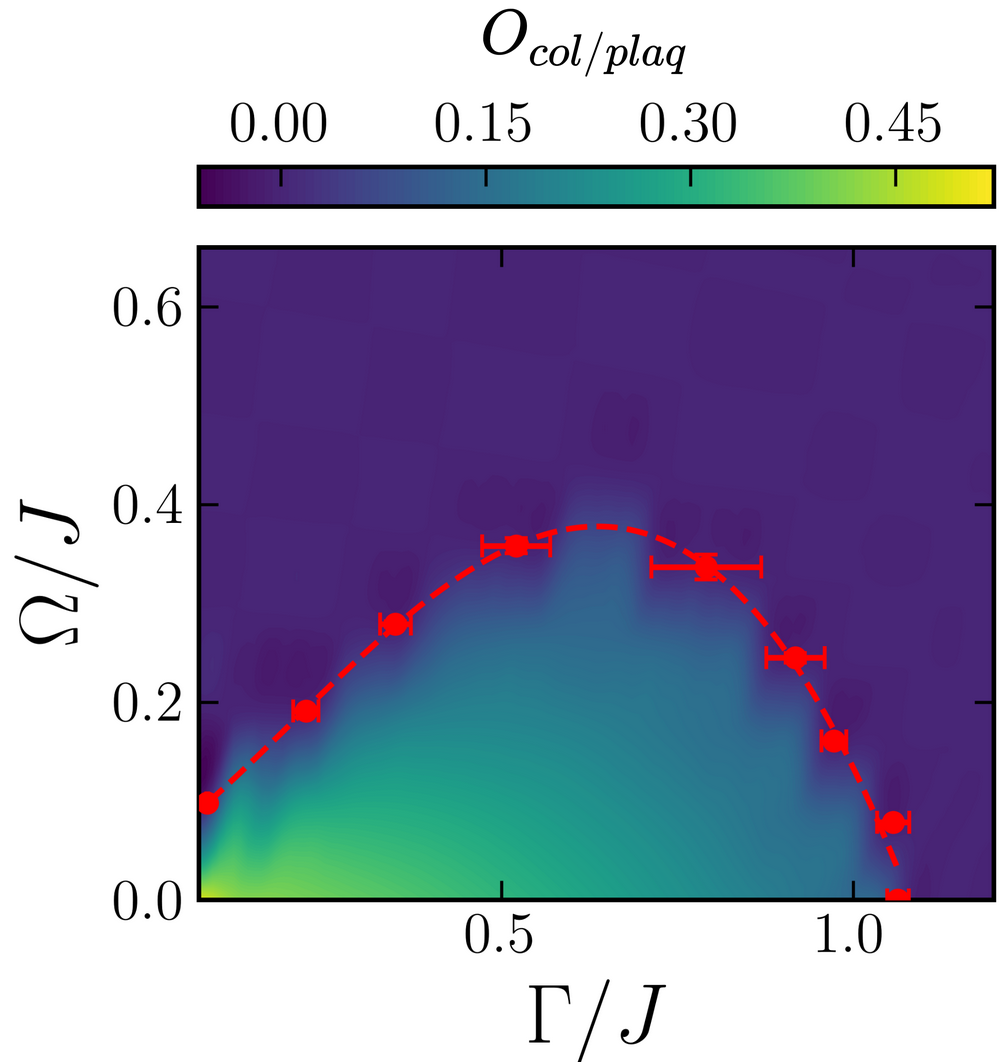}
}
\end{subfigure}
\caption{{\bf iDMRG scans of the phase diagram of Eq. \eqref{eq_gauge_invariant_H_deconf_full} in the $\Gamma/J-\Omega/J$ plane for an infinite cylinder of circumference, $L_y=4$:} (a) correlation length, $\xi$, (b)  bipartite von-Neumann entanglement entropy, $S$, (c) s-VBS order parameter $O_{stag}$ (Eq. \eqref{eq_ostag}), (d) c/p-VBS order parameter $O_{col/plaq}$ (Eq. \eqref{eq_ocol}). We take $\kappa=10$, $J=1$ and the iDMRG bond dimension, $\chi=300$. The red dots indicate the points where the c/p-VBS order parameter vanishes, while the yellow dots denote the locations of discontinuities in the s-VBS order parameter. These points are obtained from various cuts discussed below.}
    \label{fig_interpolated_order_300}
\end{figure}

We computed different observables in the ground state, including the correlation length, $\xi$, the entanglement entropy, $S$, of a bipartite cut of the cylinder, VBS order parameters, and the mean value of the star operator
\begin{align}
    O_{\rm star} = \frac{1}{N}\sum_i\langle\prod_{j\in i}X_{ij}\rangle,
    \label{eq_ostar}
\end{align}
where the summation is carried out over a unit cell, chosen to be $4\times4$ in our simulations. $N$ denotes the number of sites in the unit cell.
All VBS order parameters can be extracted from the following expression
\begin{align}
    O({\bf q}, {\bf n}) = \frac{1}{N}\sum_{{\bf r}} e^{i{\bf q}\cdot{\bf r}}\langle Z_{{\bf r},{\bf r}+{\bf n}}\rangle,
\end{align}
{where, ${\bf q}=(q_x, q_y)$ is the wave vector within the Brillouin zone and ${\bf n}$ is a real space lattice vector.} Specifically, the two components of the s-VBS order parameter and the magnitude of the order parameter are respectively defined by
\begin{align}
    O_{1(2)} = O({\bf q}&=(\pi,\pi),{\bf n}={\bf \hat x}({\bf \hat y}))
\end{align}
and
\begin{align}
    O_{\rm stag} = &\sqrt{O_1^2+O_2^2}.
    \label{eq_ostag}
\end{align}
While the c/p-VBS order parameter is given by
\begin{align}
    O_3 = O({\bf q}&=(0,\pi),{\bf n}={\bf \hat x}),~~O_4 = O({\bf q}=(\pi,0),{\bf n}={\bf \hat y})   \nonumber\\
    &O_{\rm col/plaq} = \sqrt{O_3^2+O_4^2}.
    \label{eq_ocol}
\end{align}

Two-dimensional scans of the correlation length, entanglement entropy and two VBS order parameters are illustrated in Fig. \ref{fig_interpolated_order_300}.
We clearly observe phase boundaries separating three phases sketched in Fig. \ref{fig_phase_diagram}. In particular, the staggered (columnar/plaquette) order parameter is non-zero for $\Gamma<(>)\Omega$ in the dimer limit. While the s-VBS phase persists for larger $\Omega/J$, the c/p-VBS gives way to the $\mathbb{Z}_2$ liquid at larger $\Gamma/J$. After summarizing the numerically obtained phase diagram, we now delve into the details of various parts of it.

\subsection{\texorpdfstring{$\Omega\ll \Gamma$}{}: \texorpdfstring{$\pi$}{}-flux toric code in longitudinal Zeeman field}

\begin{figure}
\includegraphics[scale=0.28]{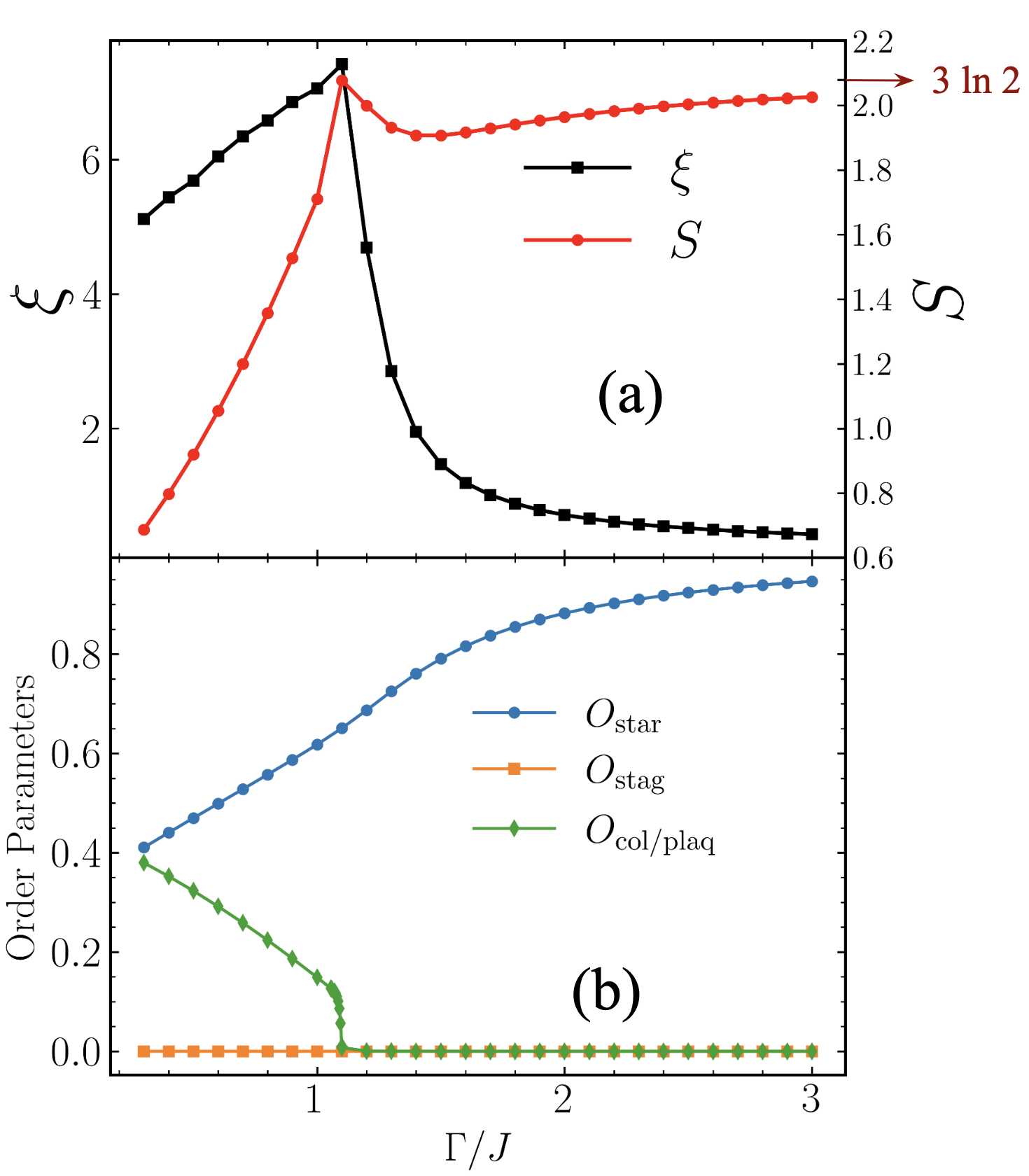}
\caption{{\bf iDMRG data along the $\Omega = 0$ line for an infinite cylinder with circumference $L_y = 4$, $\kappa = 10$ and $J = 1$ (bond dimension $\chi = 250$):} (a) Correlation length $\xi$ and bipartite von-Neuman entanglement entropy, $S$. (b) c/p-VBS order parameter $O_{\mathrm{col/plaq}}$ (Eq. \eqref{eq_ocol}), s-VBS order parameter $O_{\mathrm{stag}}$ (Eq. \eqref{eq_ostag}), and the mean value of the star operator $O_{\mathrm{star}}$.}
\label{fig_omega0}
\end{figure}

We start with the $\Omega=0$ line, where the Hamiltonian (Eq. \eqref{eq_gauge_invariant_H_deconf_full}) reduces to the $\pi$-flux toric code in the presence of a longitudinal Zeeman field. The non-trivial many-body entangled nature of the toric code phase is evident from Fig. \ref{fig_omega0}(a), where we plot the bipartite von-Neumann entanglement entropy as a function of $\Gamma/J$. Its saturation value at large $\Gamma$ approaches $(L_y-1)\log 2$~\cite{hermanns2017entanglement}, the entanglement entropy of the toric code under a bipartition that divides the infinite cylinder into two half-cylinders. In Fig. \ref{fig_omega0}(a), we also plot the correlation length, $\xi$, as a function of $\Gamma/J$. The peak signals a phase transition from the c/p-VBS to the $\mathbb{Z}_2$ topological liquid. 

Fig.~\ref{fig_omega0}(b) shows the behaviour of the VBS order parameters: the c/p-VBS order parameter acquires a non-zero value at small $\Gamma/J$ regime and vanishes at large $\Gamma/J$ in the toric code phase. Indeed, the c/p-VBS order parameter turns on at the same point as the peak of the correlation length, as is evident from the plot. The smooth change of the c/p-VBS order parameter near the phase transition indicates its possible continuous nature. The s-VBS order parameter, on the other hand, is zero throughout the parameter range.

To obtain additional insights, we also plot the expectation value of the vertex operator, $O_{star}$. Deep inside the topological liquid phase ($\Gamma\gg J $), it saturates to unity, indicating the absence of electric charges (Appendix \ref{appen_z2gt}). By decreasing $\Gamma/J$, the expectation value decreases smoothly, indicating proliferation of electric charges. Due to the non-trivial nature of the phase transition, this local expectation value does not constitute an order parameter that can diagnose the transition.  

\subsection{VBS crystals in \texorpdfstring{$\Gamma,\Omega\ll J$}{} regime: columnar/plaquette vs staggered order}

\begin{figure}
    \centering
    \subfigure[]{
        \includegraphics[width=0.22\textwidth]{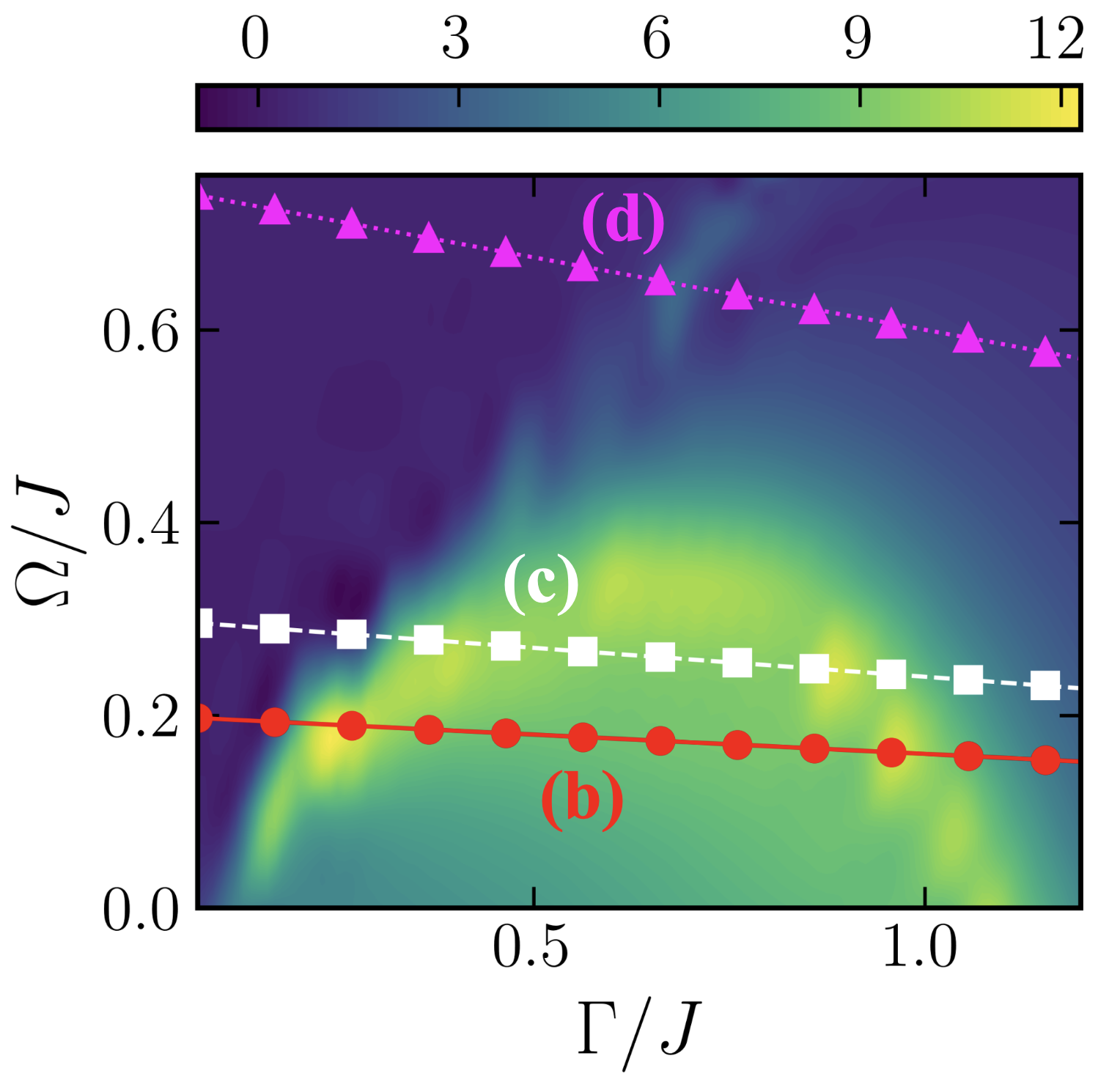}
        \label{fig:corr_length}
    }    
    \subfigure[]{
        \includegraphics[width=0.22\textwidth]{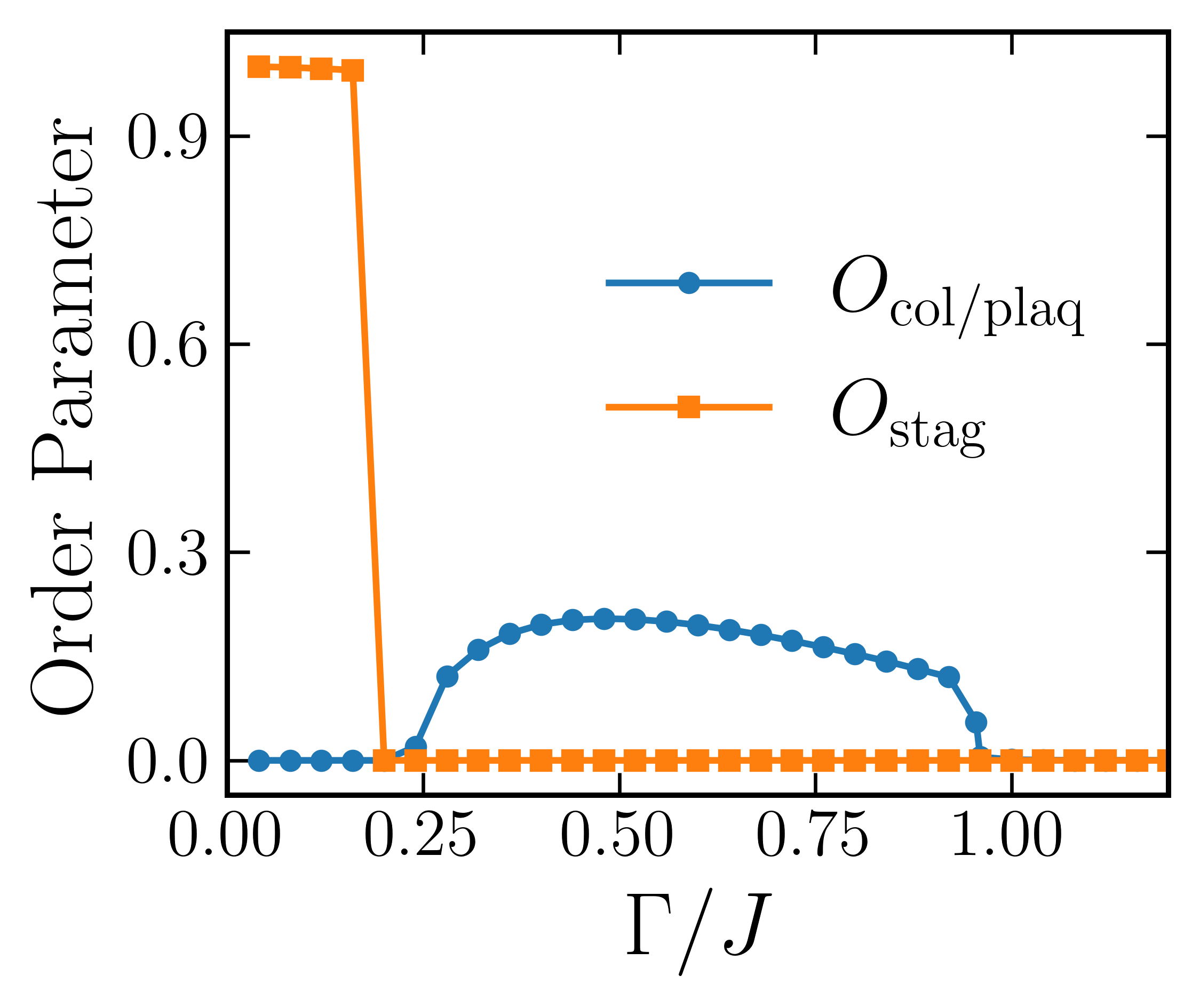}
        \label{fig:col_op}
    }
    \subfigure[]{
        \includegraphics[width=0.22\textwidth]{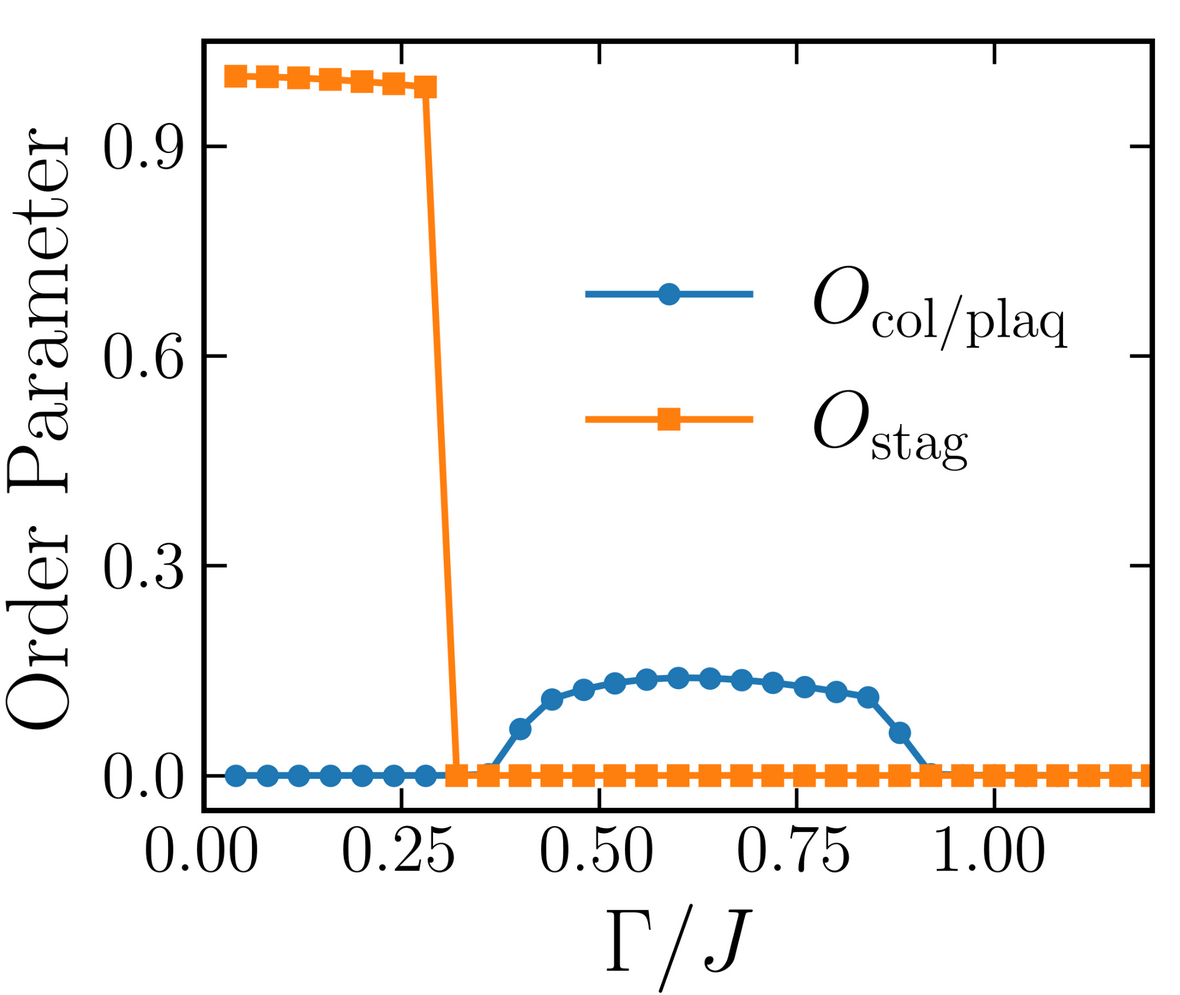}
        \label{fig:stag_op}
    }
    \subfigure[]{
        \includegraphics[width=0.22\textwidth]{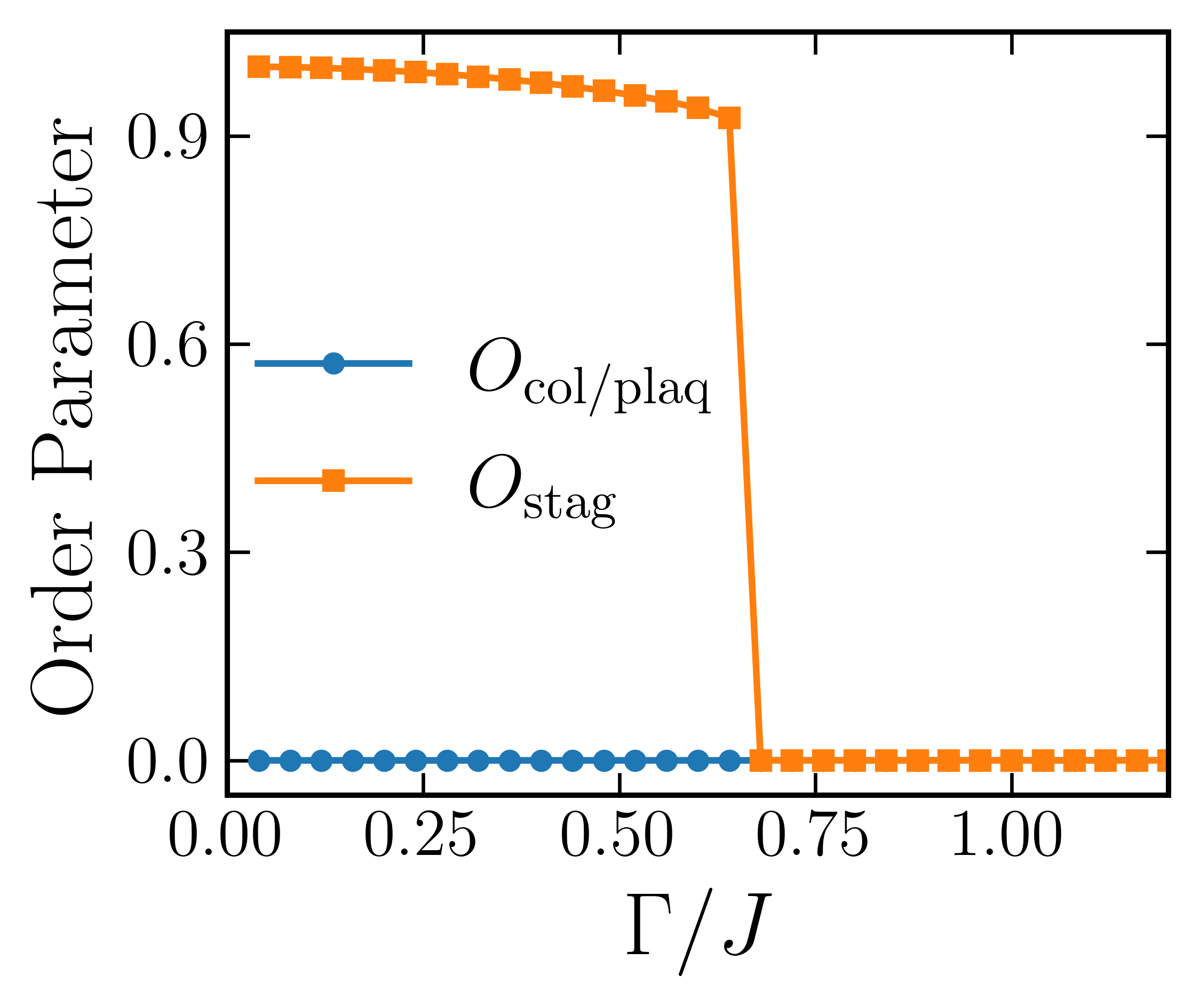}
        \label{fig:ent_entropy}
    }
    \caption{{\bf iDMRG data ($L_y=4, \kappa=10, J=1, \chi=300$) for $\Omega\neq 0$ cuts :} Columnar/plaquette and staggered order parameters (b-d) for three cuts highlighted in the contour plot (a) of the correlation length (Fig. \ref{fig_interpolated_order_300}(a)).} 
    \label{fig_interpolated_order_300_b}
\end{figure}

Now we investigate VBS orders away from the $\Omega=0$ line. Specifically, in Fig.~\ref{fig_interpolated_order_300_b}  we plot the columnar/plaquette and staggered order parameters following three cuts illustrated in panel (a). For the cuts (b) and (c), we clearly observe a transition from the staggered to columnar/plaquette order as the parameter $\Gamma/J$ is increased. While the staggered order parameter undergoes a discontinuous jump, the columnar/plaquette order parameter builds up continuously. {While this dichotomy is a hallmark of the deconfined RK criticality on the square bipartite lattice~\cite{moessner2010quantum}, we notice that there is a tiny parameter window -- noticeable in Fig. \ref{fig_interpolated_order_300_b}(b,c) (see also Appendix \ref{appen_numerical}) between the columnar and staggered phases. Because our present numerics is limited by the small cylinder circumference, we cannot definitively exclude two possibilities: One is the appearance of VBS phases with intermediate tilts, as predicted in Refs.~\cite{PhysRevB.69.224416,PhysRevB.69.224415}, which is also consistent with the field theory developed in this work (see Sec.~\ref{sec_EFT}) based on the soft electric modes of the $\mathbb{Z}_2$ liquid. The alternative interesting possibility is {the intermediate phase with $\mathbb{Z}_2$ topological order connected smoothly to the $\pi$-flux toric code.}

As one increases the slope of the cut, see panel (d) in Fig. \ref{fig_interpolated_order_300_b}, the intermediate c/p-VBS phase disappears, giving rise to a direct first-order transition from the s-VBS to the $\mathbb{Z}_2$ topological phase.

\subsection{\texorpdfstring{Phase transitions}{}}

\begin{figure}
    \centering
    \includegraphics[width=1.0\linewidth]{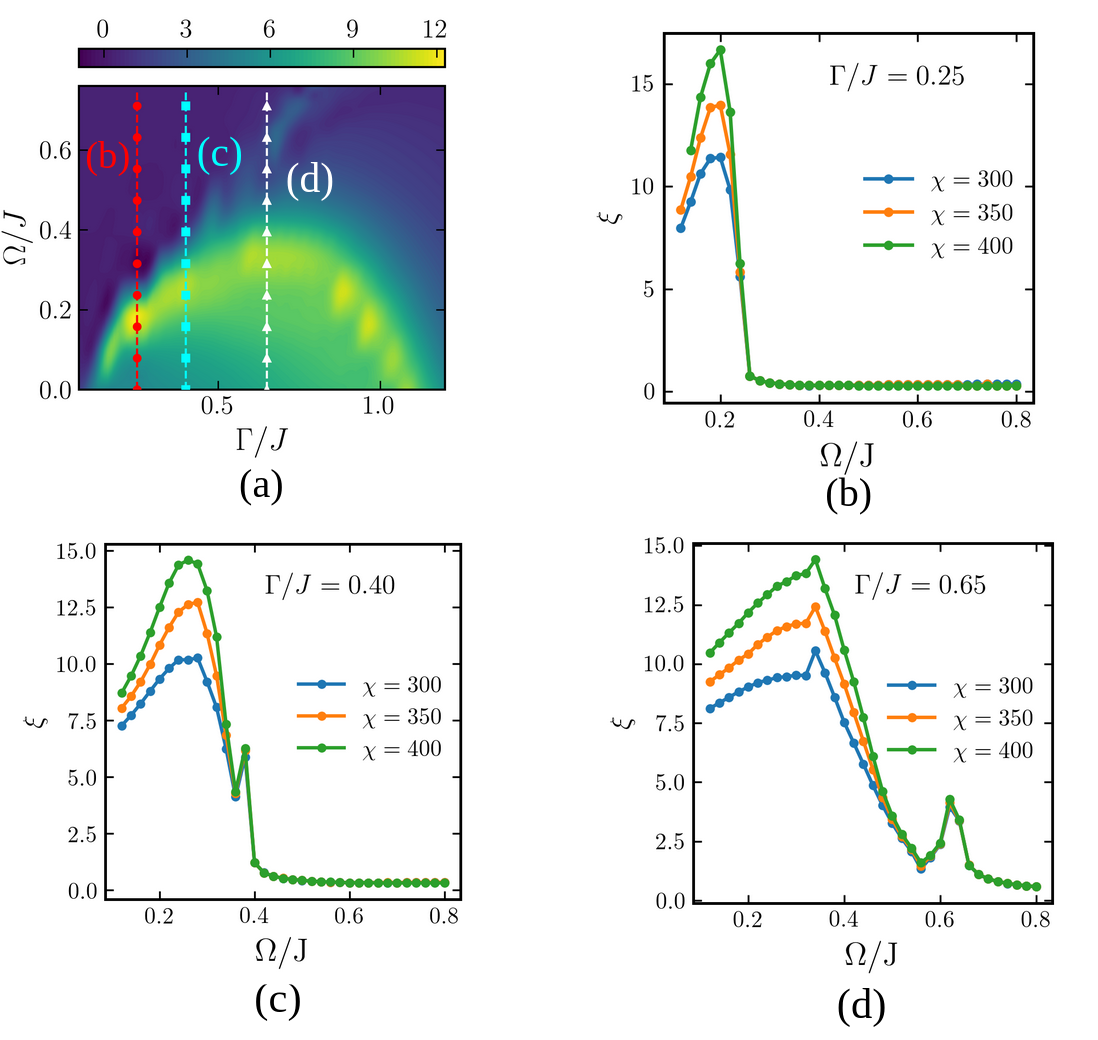}
    \caption{{\bf The location of the phase transitions:} (a) Contour plot of the correlation length obtained from iDMRG with bond dimensions $\chi=300$ on an infinite cylinder with $L_y=4, \kappa=10, J=1$. (b, c, d) Correlation length for three vertical cuts for bond dimensions $\chi=300, 350, 400$.}
    \label{corr_length_2}
\end{figure}

To obtain more information about phase transitions, we analyse the iDMRG correlation length in Fig. \ref{corr_length_2}. The cut (b) exhibits a peak whose height increases with bond dimension, $\chi$, indicating a continuous phase transition between the c/p-VBS and s-VBS phase. In panel (c), in addition to this phase transition, a new peak emerges at higher $\Omega/J$ whose moderate height does not change appreciably with increasing bond dimension. This is interpreted as the first-order phase transition separating $\mathbb{Z}_2$ topologically ordered or s-VBS phases. {Noticeably, the peak at lower $\Omega/J$, which is a continuation of the peak in (b), retains its second-order character, but broadens. In light of the discussion above, within the present resolution of our numerics, we cannot rule out a multiple-step transition between the zero tilt c/p-VBS and the maximally tilted s-VBS.} At larger values of $\Gamma/J$, one can observe in panel (d) broadening due to the $\mathbb{Z}_2$ topologically ordered phase that is separated from the c/p-VBS and s-VBS phases by a continuous and discontinuous transition, respectively ({see Appendix~\ref{appen_numerical} for more detailed analysis}).

The above iDMRG studies, while limited by system size (in the transverse direction), indicate that the {continuous quantum Lifshitz line of transition~\cite{fradkin2013field,ardonne2004topological,PhysRevB.69.224416,PhysRevB.69.224415} between the c/p-VBS and the s-VBS with possibly a thin sliver of VBS phases with intermediate tilt. This line and the line} corresponding to the continuous transition between the $\mathbb{Z}_2$ topological liquid and the c/p-VBS meet at a multicritical point and give rise to the first-order transition line between the s-VBS and the topological liquid as indicated in Fig. \ref{fig_phase_diagram}. While further numerical studies are needed to clarify the eventual fate of the phase transition lines and the multicritical point, it is useful to consider the {possible nature of the critical theory that describes the phase diagram. 

\section{The field theory for the phases and the transition} 
\label{sec_EFT}

Having obtained the numerical phase diagram, we now turn to the field theory for the phases and the phase transitions. We start deep inside the $\mathbb{Z}_2$ liquid.  As noted earlier, the Zeeman term, $J$,  gives dynamics to the electric charges, eventually leading to their condensation to produce the dimer orders. To obtain the continuum field theory, it is useful to introduce the $\mathbb{Z}_2$ electric charge field manifestly, and this can be obtained by casting the Hamiltonian \eqref{eq_gauge_invariant_H_deconf} in terms of an IGT with dynamic electric charges hopping in the background of $\pi$-flux. At $\Omega=0$, the effective Hamiltonian, after duality transformation, is given by the fully frustrated nearest neighbour Ising model in the transverse field (see Appendix \ref{appen_z2gt} for details). On the other hand, the QDM on the square lattice is best captured using a compact $U(1)$ lattice gauge theory with Eq. \eqref{eq_gaussdimer} being the electric Gauss's law~\cite{fradkin2013field,sachdev1999translational,PhysRevB.42.4568}. Such a $U(1)$ compact QED is always confining in (2+1) dimensions~\cite{polyakov81003gauge}, with the dimer crystals being the confining phases and the RK point being a fine-tuned deconfined $U(1)$ liquid~\cite{moessner2010quantum,fradkin2013field}. Below, we show that all the above aspects can be explained systematically by starting with the right (projective) symmetry allowed action for the soft electric modes supplemented with the Ising Chern-Simons term that implements the mutual semionic statistics between the electric and magnetic charges of the $\mathbb{Z}_2$ liquid.

The starting point of our analysis is to write the toric code in $\pi$-flux as an IGT coupled to electric charges~\cite{PhysRevB.102.235124} that reduces to the nearest neighbour (on the $\Omega=0$ line) fully frustrated transverse-field Ising model with the $J$ term leading to hopping of electric charges. This is done in Appendix \ref{appen_z2gt}. The ``paramagnetic'' phase of this model corresponds to the $\mathbb{Z}_2$ liquid with gapped electric (and magnetic) charges. Upon increasing  $J$, the dispersion of the electric charges develops two soft modes with amplitudes given by a pair of real fields~\cite{PhysRevB.30.1362}
\begin{align}
    (\phi_1({\bf x}), \phi_2({\bf x}))
    \label{eq_orderp}
\end{align}
at momenta $(0,0)$ and $(0,\pi)$ (in the particular gauge choice of Fig. \ref{fig_2_site_unit_cell} that implements the $\pi$-flux) and transform under a projective representation of the underlying lattice symmetries~\cite{sachdev1990effective,PhysRevB.102.235124,PhysRevB.30.1362} as detailed in Table \ref{tab_symproj} of Appendix \ref{appen_z2gt}. It is useful to combine the two real soft modes to obtain a single complex boson
\begin{align}
    \Phi=\phi_1+i\phi_2=|\Phi| e^{i\theta}.
    \label{eq_softphase}
\end{align}

As detailed in Appendix \ref{appen_qft}, by starting from these soft modes coupled with Ising gauge fields, we obtain the continuum low-energy field theory given by the 3D Euclidean action in continuum
\begin{align}
    \mathcal{S}_0=\mathcal{S}_\Phi[a]+\mathcal{S}_m[b]+\mathcal{S}_{\rm CS}[a,b]+\mathcal{S}_{\rm sym}[\Phi],
\label{eq_action}
\end{align}
where 
\begin{align}
    \mathcal{S}_\Phi[a]=&\int d^2{\bf x}d\tau\left[\frac{1}{2}|(\partial_0-ia_0)\Phi|^2+\frac{\eta}{2}|(\nabla-i{\bf a})\Phi|^2\right]\nonumber\\
    &+\int d^2{\bf x}d\tau \left[\frac{r}{2}|\Phi|^2+\frac{u}{4}|\Phi|^4\right]\nonumber\\
    & +\frac{1}{e^2}\int d^2{\bf x}d\tau~(\partial_\mu {a}_\nu-\partial_\nu a_\mu)^2
\end{align}
is the action for the electric charges coupled to the dynamic gauge field $a_\mu$. We have separated the spatial and temporal directions of the kinetic term for later convenience to access the Lifshitz transition between the different VBS phases.

Similarly, $\mathcal{S}_m[b]$ corresponds to the action for bosonic magnetic charge coupled to the dynamic $U(1)$ gauge field $b_\mu$. It has the same structure as $\mathcal{L}_\Phi$, albeit with different coupling constants. However, since the magnetic charges are always gapped in the limit of $\kappa\rightarrow\infty$ of the microscopic model, we only write the Maxwell term of the action given by
\begin{align}
    \mathcal{S}_m[b]=&\frac{1}{e^2}\int d^2{\bf x}d\tau~(\partial_\mu {b}_\nu-\partial_\nu b_\mu)^2.
\end{align}

The third piece of the action, 
\begin{align}
    \mathcal{S}_{CS}=i\frac{\epsilon^{\mu\nu\lambda}}{\pi}\int d^2{\bf x}d\tau~a_\mu\partial_\nu b_\lambda
\label{eq_cs}
\end{align}
is the mutual Chern-Simons term~\cite{PhysRevB.78.155134,freedman2004class,PhysRevB.79.064405,PhysRevB.102.235124} that imposes the mutual semionic statistics between the electric and magnetic charges. As we shall see below, in the rest of this discussion, the effect of the CS term is mostly to fix the right topological degeneracy in the $\mathbb{Z}_2$ liquid state. 

Finally, $\mathcal{S}_{\rm sym}[\Phi]$ denotes the microscopic symmetry allowed terms under which the soft modes are invariant (see Table \ref{tab_symproj}). On the square lattice, the lowest order contribution is given by~\cite{PhysRevB.30.1362}
\begin{align}
    \mathcal{S}_{\rm sym}[\Phi]=-w\int d^2{\bf x}d\tau~|\Phi|^8\cos(8\theta)
\end{align}
which destroys the electric charge conservation and hence reduces the symmetry of the above action from $O(2)$ to $\mathbb{Z}_8$. Clearly, this term allows for the simultaneous creation/annihilation of {\it eight} electric charges. Noticeably, the CS term, capturing the semionic statistics, shows that such an event corresponds to the change of flux of $b_\mu$ by 4 units and hence corresponds to quadrupoled instanton events for $b_\mu$~\cite{sachdev2018topological}. Indeed such quadrupoled monopole creation/annihilation operators are allowed~\cite{polyakov81003gauge} by the symmetry of the square lattice~\cite{PhysRevB.42.4568,PhysRevLett.61.1029,sachdev1990effective,fradkin2013field} and $\mathcal{S}_{\rm sym}[\Phi]$ can alternatively be written as
\begin{align}
    \mathcal{S}_{\rm sym}[\Phi]\sim -w\int d^2{\bf x}d\tau~\left(\mathcal{M}_b^4+{\rm c.c.}\right),
\end{align}
where $\mathcal{M}_b$ annihilates a bosonic monopole of $b_\mu$ with the identification 
\begin{align}
    \mathcal{M}_b\sim e^{i2\theta}
    \label{eq_monopoleheight}
\end{align}
where the symbol $\sim$ refers to the fact that the two operators have the same symmetry quantum numbers and scaling dimension~\cite{sachdev1990effective,PhysRevB.42.4568}.

Once $\Phi$ condenses, the sign of $w$ chooses the columnar ($w>0$) or plaquette ($w<0$) order via the order parameters 
\begin{align}
    (O_3, O_4)\sim (\phi_1^2-\phi_2^2, 2\phi_1\phi_2) =|\Phi|^2(\cos2\theta, \sin 2\theta) 
    \label{eq_colplaop}
\end{align}
This leads to the identification of the monopole operator  \eqref{eq_monopoleheight} with the exponential of the height field representation of the dimer model~\cite{fradkin2013field,sachdev1999translational,PhysRevB.42.4568,PhysRevLett.62.1694}. Note that the same continuum action  \eqref{eq_action} can be obtained via a mutual $U(1)$ Chern-Simons theory formulation~\cite{PhysRevB.79.064405,PhysRevB.102.235124}.

To complete the effective theory, the above action \eqref{eq_action} needs to be supplemented with a {\it Lifshitz} term given by the Lagrangian
\begin{align}
    \mathcal{S}_l[\Phi]=&\frac{\eta_2}{2}\int d^2{\bf x}d\tau~|(\nabla-i{\bf a})^2\Phi|^2\nonumber\\
    &+\frac{\lambda_2}{2}\int d^2{\bf x}d\tau~\left\{[(\partial_1-ia_1)^2\Phi^*][(\partial_2-ia_2)^2\Phi]\right.\nonumber\\
    &~~~~~~~~~~~~~~~~~~~~~~~~~~~~~~~~~~~~~~~~~~~~~~~~~~~\left.+~{\rm c.c.}\right\}
    \label{eq_lifshitz}
\end{align}

with $\eta_2,\lambda_2(>0)$ are the coupling constants of the two terms allowed by the PSG. Therefore, the total action is given by
\begin{align}
    \mathcal{S}=\mathcal{S}_0+\mathcal{S}_l
    \label{eq_finalaction}
\end{align}

This allows us to incorporate the Lifshitz transition when $\eta \sim \Gamma-\Omega$~\cite{fradkin2013field} changes sign, as we explain below. We now analyse the mean field phase diagram of this action that results in Fig. \ref{fig_fpd}.

\subsection{Phases and phase transitions}

In the above field theory (Eq. \eqref{eq_finalaction}), the electric soft modes are gapped for $r>0$ and can be integrated out, resulting in the pure $U(1)\times U(1)$ CS theory. This is nothing but the $\mathbb{Z}_2$ liquid with a 4-fold degenerate ground state on a torus. Alternatively, as shown in Appendix \ref{appen_qft}, the action, $\mathcal{S}_0$, in Eq. \eqref{eq_action} without the monopole terms is dual to a charge-2 Higgs scalar coupled to the $U(1)$ gauge field (Eq. \eqref{eq_abelianhiggs2e}). In this dual description, the $\mathbb{Z}_2$ liquid is obtained via condensation of the charge-2 Higgs scalar, whence the gauge group reduces to $\mathbb{Z}_2$~\cite{PhysRevB.44.686,sachdev1999translational} and the topological liquid represents the deconfined phase of this IGT. Within the mean field, this $\mathbb{Z}_2$ liquid is obtained for $r,\eta>0$ as shown in the phase diagram in Fig. \ref{fig_fpd}.

\begin{figure}
    \centering
    \includegraphics[width=0.95\linewidth]{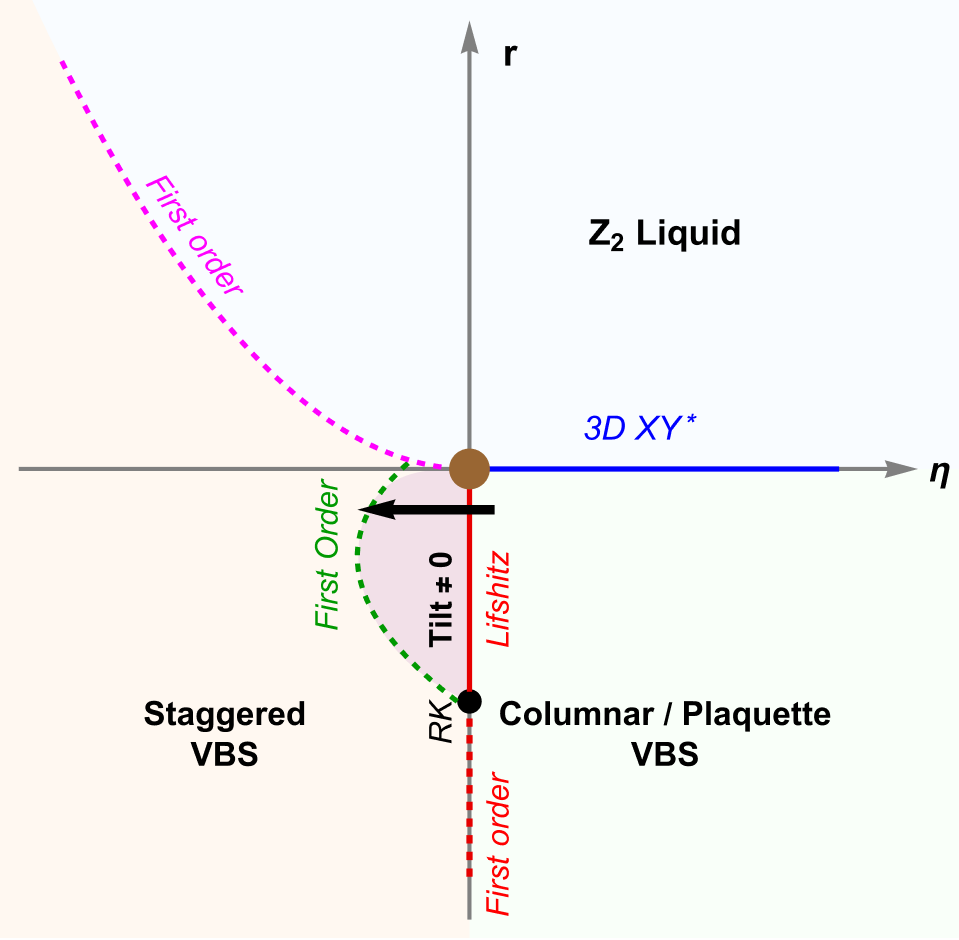}
    \caption{{\bf The mean field phase diagram obtained from the low-energy field theory (Eq. \eqref{eq_finalaction}):}  The $\mathbb{Z}_2$ liquid dominates the region where $\eta,r>0$, whence the soft electric modes are gapped. The zero tilt c/p-VBS is obtained by condensing the electric soft modes ($r<0$) for $\eta>0$ via a 3D XY$^*$ transition (blue solid line). The s-VBS, on the other hand, is obtained for $\eta,r\ll 0$. On crossing the $\eta=0$ line from the $\eta>0$ side, there is a transition between the zero tilt c/p-VBS to the maximally tilted s-VBS phase. This transition can happen in two ways~\cite{PhysRevB.69.224415,PhysRevB.69.224416} as indicated -- The red dotted line is a first-order transition between the c/p-VBS and the s-VBS phases while the red solid line denotes a continuous quantum Lifshitz transition between the c/p-VBS and VBS phase with finite tilt (The black arrow indicates the expected increase in the tilt~\cite{PhysRevB.69.224415,PhysRevB.69.224416}) with the RK point being the multicritical point in between. Various other first-order phase transitions are denoted in dashed lines as indicated -- {\it e.g.}, the transition between the $\mathbb{Z}_2$ liquid and the s-VBS is given by Eq. \eqref{eq_stagz2}. The brown point at the origin $(\eta,r)=(0,0)$ is the multicritical point where the two continuous transition lines -- Lifshitz line, the 3D XY$^*$ line -- meet to give rise to a first-order transition line between the $\mathbb{Z}_2$ liquid and the s-VBS phase. This phase diagram should be compared with the microscopic model phase diagram in Fig. \ref{fig_phase_diagram}.}
    \label{fig_fpd}
\end{figure}

Tuning $r$ from positive to negative, the electric charges condense, leading to the symmetry breaking. For $\eta>0$, the higher order gradient term  in Eq. \eqref{eq_lifshitz} are irrelevant and thus $\eta_2, \gamma$ and $\lambda$ can be set to zero. Therefore, the transition driven by $r$, in this regime, is given by the Abelian Higgs model with 8-fold anisotropy due to the $w$ term. However, the anisotropy is dangerously irrelevant at the transition~\cite{PhysRevB.30.1362} and the critical theory is given by an Abelian Higgs model and belongs to the 3D XY$^*$ universality class as indicated in Fig. \ref{fig_fpd}. The dangerously irrelevant term, however, chooses the columnar (plaquette) VBS for $w>0$ as signaled by the order parameter in Eq. \eqref{eq_colplaop}. 

{Sitting inside the zero tilt c/p-VBS (for $r<0, \eta>0$), it is useful to consider tuning $\eta$ across zero to a negative value. This leads to the transition between the zero tilt c/p-VBS phase (for $\eta\gg 0$) to the maximally tilted s-VBS (for $\eta\ll 0$). However the tilt can change continuously and this, as we show now, in general mediates an intricate multi-step transition, as indicated in Fig. \ref{fig_fpd}.}

Deep inside the Higgs phase ($r\ll 0$) the amplitude fluctuations, $|\Phi|$, are frozen leading to the mean field action 
\begin{align}
    \mathcal{S}_{\rm eff}[r\ll 0]=&\frac{1}{2}\int d^2{\bf x}d\tau~\left[\eta{\bf a}^2+\eta_2{\bf a}^4+2\lambda_2 a_1^2a_2^2\right]|\Phi|^2\nonumber\\
    &+\mathcal{S}_{\rm sym}[\Phi]
    \label{eq_seff}
\end{align}
where we have used the gauge transformation
\begin{align}
    a_\mu\rightarrow a_\mu +\partial_\mu\theta
\end{align}
and the photon of $a$ is gapped out via the usual Higgs mechanism. Similarly, because the monopoles of $b_\mu$ are condensed (Eq. \eqref{eq_monopoleheight}), the corresponding photon is gapped due to the dual Higgs mechanism (confinement). For $w=0$, the minimum of the above effective action  (Eq. \eqref{eq_seff}) is obtained for
\begin{align}
    \bar{a}_1=\pm \bar{a}_2
\end{align}
with 
\begin{align}
    a_2=\left\{\begin{array}{ll}
    0 & \eta>0\\
    \sqrt{\frac{|\eta|}{4\eta_2+2\lambda_2}} & \eta<0\\
    \end{array}\right.
    \label{eq_finitetilt}
\end{align}
Note that the CS term goes to zero for a constant $a_\mu$.

The effect of $w\neq 0$ is obtained by noting that in the Higgs phase, $a_\mu$ is a pure gauge due to the Meissner effect. We then have
\begin{align}
    \bar{a}_\mu=\partial_\mu \theta~~~~~\Rightarrow~~~~\theta({\bf x})=\bar{\bf a}\cdot {\bf x}+\theta_0
    \label{eq_higgsfluctuation}
\end{align}
such that the minima of the phase is chosen via $\mathcal{S}_{\rm sym}[\Phi]$ through the minimization of
\begin{align}
    \mathcal{S}_{\rm sym}[\Phi]=-w|\Phi|^8\int d^2{\bf x}d\tau~\cos(8\bar{\bf a}\cdot{\bf x}+8\theta_0).
\end{align}

Therefore, for $\eta>0$ the minima are given by $\bar{\bf a}=0$ giving rise to a zero tilt sector, {\it i.e.}, c/p-VBS order as given by Eq. \eqref{eq_colplaop}. On the other hand, for $\eta<0$ we get $\bar{\bf a}\neq 0$ and a finite tilt sector is chosen~\cite{PhysRevB.69.224416} in the field theory with the tilt related to the expectation value of $\bar{\bf a}$. 

The critical theory for the transition between the zero tilt and the finite tilt VBS phases in this regime ($r\ll 0$) can be obtained by restoring the low energy phase fluctuations in the above mean field action via Eq. \eqref{eq_higgsfluctuation}. The effective low-energy theory is given by
\begin{align}
    \tilde{\mathcal{S}}=\tilde{\mathcal{S}}_0+\tilde{\mathcal{S}}_l
    \label{eq_actionrl0}
\end{align}
with
\begin{align}
    \tilde{\mathcal{S}}_0=&\frac{1}{2}\int d^2{\bf x}d\tau\left[(\partial_0\theta)^2+\eta(\nabla\theta)^2-w|\Phi|^6\cos(8\theta)\right]
\end{align}
and
\begin{align}
    \tilde{\mathcal{S}}_l=&\frac{1}{2}\int d^2{\bf x}d\tau\left[\eta_2\left(\nabla^2\theta\right)^2+\lambda_2(\partial_1^2\theta)(\partial_2^2\theta)\right]\nonumber\\
    &+\frac{1}{2}\int d^2{\bf x}d\tau~\left[\eta_2(\nabla\theta)^4+\lambda_2(\partial_1\theta)^2(\partial_2\theta)^2\right].
\end{align}
where we have absorbed the overall $|\Phi|^2$ in the coupling constants. This is precisely the action obtained from the height model representation with the height field being proportional to $2\theta$~\cite{fradkin2013field,PhysRevB.69.224416,ardonne2004topological,henley2004classical,PhysRevB.69.224416,PhysRevB.69.224415}. For $\eta\neq 0$, the field theory is always in the confining phase~\cite{polyakov81003gauge,PhysRevB.69.224416,PhysRevB.69.224415,ardonne2004topological,fradkin2013field} while there is a transition for $\eta=0$ albeit with renormalisation due to interactions.

For $\eta=0$, the dynamic critical exponent is  $z=2$~\cite{PhysRevB.69.224416,PhysRevB.69.224415,ardonne2004topological,fradkin2013field} for the Gaussian theory, and the monopoles can become irrelevant. The irrelevance of the monopoles can be obtained from the scaling dimension of $e^{i\theta}$ for the free Lifshitz field theory gotten by collecting the Gaussian terms of Eq. \eqref{eq_actionrl0} for $\eta=0$, {\it i.e.},
\begin{align}
    \tilde{\mathcal{S}}_g=&\frac{1}{2}\int d^2{\bf x}d\tau\left[(\partial_0\theta)^2+\eta_2\left(\nabla^2\theta\right)^2+\lambda_2(\partial_1^2\theta)(\partial_2^2\theta)\right]
    \label{eq_gaussianlifshitz}
\end{align}

The equal time correlator in the presence of the anisotropy, $\lambda_2$, to the leading order, is given by (details in Appendix \ref{appen_correlator})
\begin{align}
    \langle e^{i n\theta_{\bf x}}e^{-in\theta_0}\rangle&\sim \frac{1}{|{\bf x}|^{n^2\alpha}}
    \label{eq_monopolecorrelator}
\end{align}
where $n\in \mathbb{Z}$ and 
\begin{align}
    \alpha=\frac{1}{8\pi^2\sqrt{\eta_2}}\frac{4}{\sqrt{1+\epsilon/4}}\mathcal{K}\left(\frac{\epsilon}{4+\epsilon}\right)=\frac{f(\epsilon)}{8\pi^2\sqrt{\eta_2}}
\end{align}
is the effective coupling constant with $\epsilon=\lambda_2/\eta_2$. Now for the $w$ term to be irrelevant~\cite{PhysRevB.69.224416}, we want $\alpha>\frac{1}{8}$, or,
\begin{align}
    \frac{f(\epsilon)}{\sqrt{\eta_2}}>\pi^2~~~~\Rightarrow~~~~~~0< \eta_2<\frac{f^2(\epsilon)}{\pi^2}
\end{align}
where $f(0)=2\pi$ and decreases monotonically with $\epsilon$. It is therefore interesting to note that the window of coupling constants where the quadrupoled monopoles are irrelevant decreases as $\lambda_2/\eta_2$ increases. More specifically 
\begin{align}
    \lim_{\epsilon\rightarrow\infty} f(\epsilon)\sim \frac{8}{\sqrt{\epsilon}}\ln\epsilon
\end{align}

Thus, the regime over which the monopoles are irrelevant shrinks to zero as $\epsilon\rightarrow\infty$. Therefore, the increase in the anisotropy decreases the regime over which the Lifshitz criticality is stable to the monopole proliferation. This is due to the decrease in the scaling dimension of $e^{i\theta}$ with increasing anisotropy. When such monopoles are relevant, it leads to a first-order transition between the c/p-VBS (for $\eta>0$) to s-VBS (for $\eta<0$).

Within the regime where the monopoles are irrelevant, the effect of the interaction, given by (from Eq. \eqref{eq_actionrl0}),
\begin{align}
    \mathcal{S}_{\rm int}
    &=\int d^2{\bf x}d\tau\left[\frac{\eta_2}{2}\left\{(\partial_1\theta)^4+(\partial_2\theta)^4\right\}\right.\nonumber\\
    &~~~~~~~~~~~~~~~~~~~~~~~~~\left.+\frac{\lambda_2+6\eta_2}{2}(\partial_1\theta)^2(\partial_2\theta)^2\right]
\end{align}
can be considered with perturbative RG. Indeed Ref. \onlinecite{PhysRevB.69.224416,PhysRevB.69.224415} calculated the RG flow of the two coupling constants to one loop to indicate that except for further fine tuning of the ratio of the two interactions, the Gaussian critical theory is unstable and there is a generic first-order transition~\cite{aubry1983discrete} between the c/p-VBS phase and the s-VBS phase that ends at the RK point. Beyond the RK point, the columnar/plaquette order gives way to a finite (non-maximal) tilt phase via a continuous transition at the quantum Lifshitz line characterised by $\eta=0$-- albeit with renormalisation. Beyond this, the tilt changes via an incomplete devil staircase, ultimately giving way to the s-VBS phase via a first-order phase transition. This is indicated for the $r\ll0$ regime of the Fig. \ref{fig_fpd}. 

The above conclusion based on the phase fluctuation is strictly valid when $r\ll 0$ and hence breaks down on approaching $r\sim 0$ when the amplitude fluctuations of the soft electric modes can no longer be neglected. This can lead to a major departure from the conclusion of the above analysis based on Ref. \onlinecite{PhysRevB.69.224416,PhysRevB.69.224415}. However, before we turn to this issue, let us consider the quadrant with $r>0$ but $\eta<0$. For this, we consider the position of the minima of the quadratic part of the action (Eq. \eqref{eq_action} and \ref{eq_lifshitz}). 

Neglecting the gauge fluctuations and the temporal fluctuations of $\Phi$, the mean field free energy in momentum space is given by
\begin{align}
    \frac{1}{2}\int d^2{\bf q}~\left[\eta {\bf q}^2+\eta_2{\bf q}^4+r\right]|\Phi({\bf q})|^2.
\end{align}
Such that the momentum of the soft modes, ${\bf q}_0(=\bar{\bf a})$, is given by Eq. \eqref{eq_finitetilt}. These soft modes condense, for $\eta<0$ for 
\begin{align}
    r\sim \frac{\eta^2}{\eta_2}
    \label{eq_stagz2}
\end{align}
as given by the dashed line in Fig. \ref{fig_fpd}. Thus, the $\mathbb{Z}_2$ liquid gives way to a condensate of the electric charges at ${\bf q}_0$ -- a VBS phase -- on crossing this line. Due to the fluctuations near the transition, we expect this transition to be first-order.

The above first-order line then ends at $(\eta,r)=(0,0)$ with the 3D XY$^*$ continuous transition line as well as the Lifshitz transition line between the c/p-VBS and the finite tilt phase as indicated in the Fig. \ref{fig_fpd}. In principle, the finite tilt phase can extend above the $r=0$ line, albeit for $r<\frac{\eta^2}{\eta_2}$, and surround the multicritical point, but our numerical calculations (also see the discussion below) indicate that such a region is very small as schematically indicated in the figure. The multicritical point at $(\eta,r)=(0,0)$ is therefore obtained by a $z=2$ anisotropic Abelian Higgs theory given by Eq. \eqref{eq_action} and \eqref{eq_lifshitz}.

\subsection{Applicability of the field theory to the microscopic model}

Having fleshed out the details of the field theory, we now briefly summarise the applicability of the field theory (Fig.~\ref{fig_fpd}) to the microscopic model (Eq.~\eqref{eq_gauge_invariant_H_deconf_full}) and the resultant phase diagram (Fig. \ref{fig_phase_diagram}) in the light of the numerical results discussed in Sec. \ref{sec_numerics}.

We can tune the microscopic model~\eqref{eq_gauge_invariant_H_deconf_full} to the RK point by approaching the origin in the phase diagram shown in Fig.~\ref{fig_phase_diagram} along the line $\Gamma/J=\Omega/J$. This can be achieved by tuning $J\to\infty$ while keeping $\Gamma$ and $\Omega$ finite, since the higher-order terms generated in perturbation theory, of order $\mathcal{O}(\Gamma^2/J)$ or higher, vanish in this limit. As a result, we are left with the RK Hamiltonian given in Eq.~\eqref{eq_RK_hamiltonian} at $\Gamma=\Omega$. This leads us to the identification of the RK point of Fig. \ref{fig_fpd} with the origin of Fig. \ref{fig_phase_diagram}.

Moving away from the RK point, the Lifshitz transition line emerges, as shown in Fig.~\ref{fig_fpd}. This transition is manifested in Fig.~\ref{fig_interpolated_order_300_b}(b,c) by a continuous increase of the c/p-VBS order parameter. Further, as commented in Sec. \ref{sec_numerics}, the thin sliver where both the order parameters are zero may allow for an intermediate finite tilt phase in accordance with the field theory (see Fig.~\ref{fig_fpd}).  This is also very indirectly supported by the broad correlation length peak for the corresponding window as shown in Fig. \ref{corr_length_2}. While the scope of the present numerics is limited, the above observations are prima facie consistent with the prediction of the field theory presented above. Indeed, a soft mode analysis of the electric modes at finite but small $\Omega/J$ (see Appendix \ref{appen_sfotmode_omega}) suggests that the position of the soft modes changes continuously away from the c/p-VBS for finite tilt VBS phases within Gaussian approximation. {Upon further decreasing $J$, the Lifshitz transition line meets both the first-order transition from the $\mathbb{Z}_2$ liquid to the s-VBS phase and the 3D XY$^*$ transition from the c/p-VBS phase to the $\mathbb{Z}_2$ liquid at a multicritical point (see Fig.~\ref{fig_phase_diagram}). This multicritical point corresponds to the origin in Fig.~\ref{fig_fpd}.}

Finally, beyond the multicritical point, the transition between the $\mathbb{Z}_2$ liquid and the s-VBS appears to be strongly first-order, and the tilt almost immediately locks to the maximal value (corresponding to the s-VBS phase) as is evident from Fig. \ref{corr_length_2}(d) and the discussion therein. 

\section{Summary and Outlook} \label{sec_summary}

{In this work, we have studied a particular qubit regularisation of the QDM on the square lattice that continuously interpolates the physics of the dimer models with that of the $\mathbb{Z}_2$ topological liquid as realised in the toric code with $\pi$-flux. This qubit regularisation allows for mixing the one-dimer and three-dimer manifolds with a controlled energy cost. The emergent $\mathbb{Z}_2$ liquid is intimately connected to this mixing that allows for condensation of a charge-2 Higgs scalar in the $U(1)$ theory, usually realised as the effective description of the dimer model on a bipartite lattice~\cite{sachdev1999translational,PhysRevB.42.4568,sachdev1990effective}. The resultant phase diagram illustrated in Fig. \ref{fig_phase_diagram} appears to be very rich and in particular realises an extended $\mathbb{Z}_2$ liquid phase and its transition to the dimer crystal via interesting quantum phase transitions.} 

Recent experimental works have successfully realised the toric code and diagnosed its associated $\mathbb{Z}_2$ topological order across diverse quantum platforms~\cite{weimer2011digital,weimer2010rydberg,PhysRevX.4.041037,PhysRevX.10.021057}. That includes the realisation of the toric code ground state and the demonstration of anyon braiding statistics on superconducting processors~\cite{satzinger2021realizing}, the demonstration of topological $\mathbb{Z}_2$ spin liquid in Rydberg atom arrays \cite{semeghini2021probing}, and the deterministic preparation of the toric code on a torus using mid-circuit measurements and feed-forward on a trapped-ion quantum computer ~\cite{iqbal2024topological}. Our qubit regularisation of the quantum dimer model opens a path towards its realisation with current quantum technologies that could shed additional light on how the fine-tuned RK $U(1)$ spin liquid transitions into the $\mathbb{Z}_2$ topological liquid quantum phase.

{Central to the structure of the phase diagram is a multicritical $U(1)$ liquid with dynamic critical exponent, $z=2$, that emanates} from the RK point. Various phases are naturally understood as instabilities of this $U(1)$ liquid. It would be interesting to investigate in detail the fate of the multicritical point and nearby phases and phase transitions on square and other two-dimensional lattices.}

\acknowledgments

The authors thank Kedar Damle, Souvik Kundu, Sounak Biswas and Sthitadhi Roy for useful discussions. The authors acknowledge STINT (Stiftelsen för internationalisering av högre utbildning och forskning) for funding via the Internationalisation Initiation grant. AC and SB thank Karlstad University for hospitality, and the Department of Atomic Energy, Government of India, under project no. RTI4019 and RTI4013. SM acknowledges ICTS for hospitality. The authors would like to thank the ICTS program - Generalised symmetries and anomalies in quantum phases of matter 2026 (code: ICTS/ GSYQM2026/01). SM is supported by Vetenskapsr{\aa}det (2021-03685) and Nordita. AC and SM thank Carl Tryggers Stiftelse (CTS 24:3607) for funding their research. SB gratefully acknowledges funding by the Swarna Jayanti fellowship of SERB-DST (India) Grant No. SB/SJF/2021-22/12; DST, Government of India (Nano mission), under Project No. DST/NM/TUE/QM-10/2019 (C)/7. The numerical simulations were performed on the BOSON-1 workstation at ICTS.

\appendix

\begin{figure*}
    \centering
    \includegraphics[width=1.85\columnwidth]{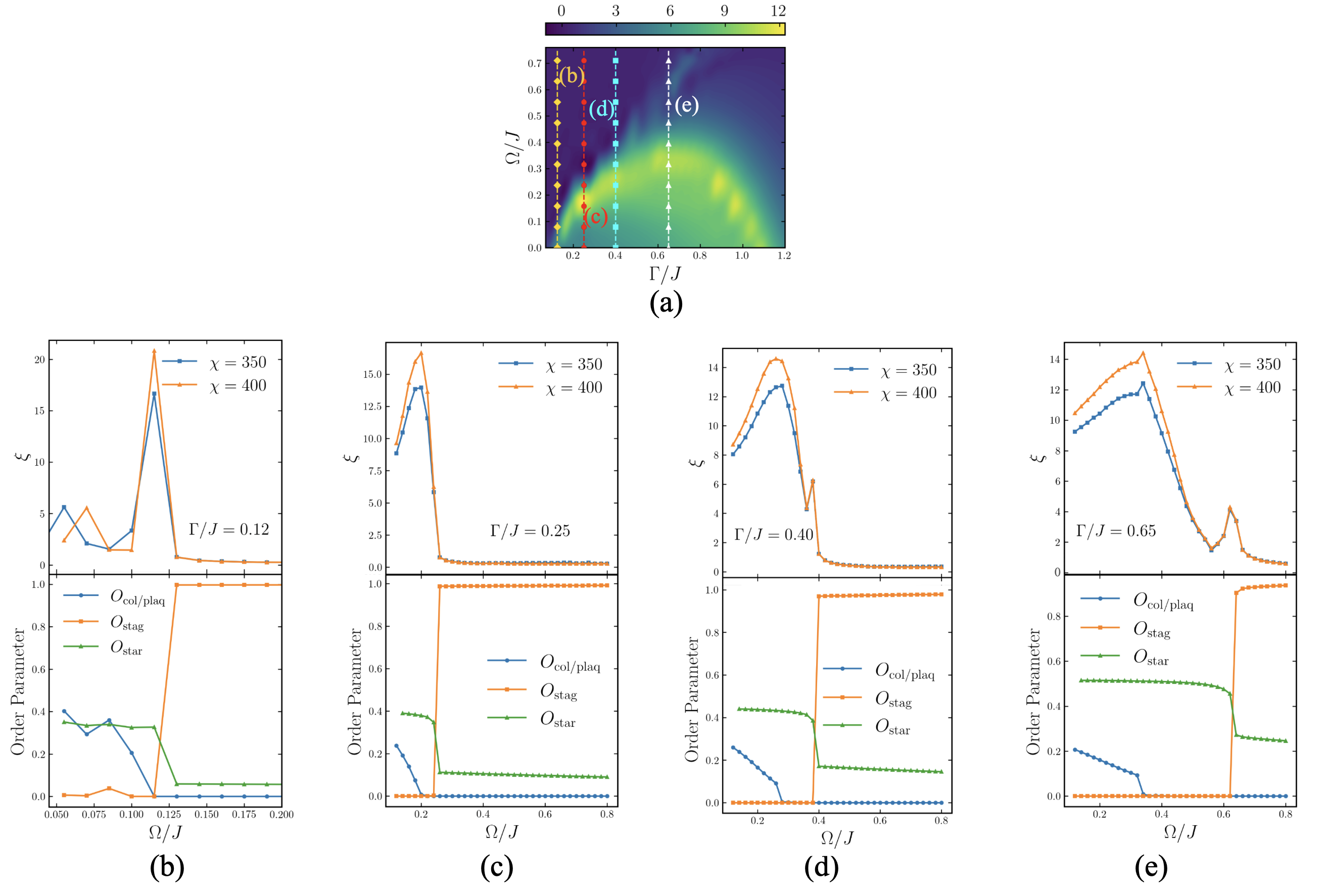}
   \caption{{\bf Vertical cuts for the iDMRG data for $L_y=4, \chi=300, \kappa=10, J=1$ :} (a) Contour plot of the correlation length (same as Fig. \ref{fig_interpolated_order_300}(a)). (b, c, d, e) Correlation length $\xi$ for two different bond dimensions and order parameters at $\chi = 400$ along the vertical cuts shown in panel (a).}
    \label{corr_ord_2}
\end{figure*}
\section{Further numerical data}
\label{appen_numerical}

In this appendix, we present further numerical details of the iDMRG calculations. In particular, we provide the plots for the two order parameters, $O_{col/plaq}$ (Eq. \eqref{eq_ocol}) and $O_{stag}$ (Eq. \eqref{eq_ostag}), along with $O_{star}$ evaluated along the vertical cuts, including those shown in Fig.~\ref{corr_length_2} of the main text. 

As illustrated in the lower panels of Fig.~\ref{corr_ord_2}, for small values of $\Gamma/J$, the system undergoes a direct transition from the c/p-VBS phase to the s-VBS phase as a function of $\Omega/J$, as evident from Fig.~\ref{corr_ord_2}(b). {At large $\Gamma/J$ corresponding to panel (d) and (e), an intermediate window corresponding to a $\mathbb{Z}_2$ liquid emerges where both order parameters vanish}  leading to a sequence of transitions from the c/p-VBS phase to the $\mathbb{Z}_2$ liquid and subsequently to the staggered phase. The former transition corresponds to the 3D XY$^*$, while the latter is a first-order transition from the $\mathbb{Z}_2$ liquid to the s-VBS phase, as discussed in the main text. {At the first-order transition from the $\mathbb{Z}_2$ liquid to the staggered phase, $O_{\text{star}}$ shows an abrupt drop, which arises from the absence of local dimer dynamics in the staggered phase.} For completeness, we also plot the correlation length $\xi$ together with the order parameters in Fig.~\ref{corr_ord_2}.  It is clear that the phase transition points identified from the order parameters are consistent with those obtained from the correlation length. {At intermediate $\Gamma/J$ corresponding to panel (c), the peak corresponding to the first order transition between the $\mathbb{Z}_2$ liquid and the s-VBS is no longer visible, but there is a small region where both $O_{stag}$ and $O_{col/plaq}$ are zero. Our present numerics cannot determine whether this region survives in the thermodynamic limit of the two dimensional lattice.}

\section{The Ising gauge theory formulation}
\label{appen_z2gt}

Following Refs. [\onlinecite{tupitsyn2010topological, PhysRevB.102.235124,PhysRevLett.98.070602,PhysRevB.91.134419}], the Hamiltonian in Eq.~\eqref{eq_gauge_invariant_H_deconf} can be mapped to a model of IGT with Ising electric charges as
\begin{align}
    Z_{ij}=\tau^x_i\rho^z_{ij}\tau^x_j,~~~~~~~X_{ij}=\rho_{ij}^x,
\end{align}
where $\tau^x_i$ creates Ising (electric) charges on the sites and $\rho_{ij}^z$ are the Ising gauge fields (hence $\rho_{ij}^x$ is the electric field). The above redundant representation is subject to Gauss's law constrained on every site
\begin{align}
    \prod_{i\in i}\rho^x_{ij}=\tau_i^z,
    \label{eq_gaussising3}
\end{align}
where $(1-\tau^z_i)/2$ is the electric charge density. Using this, the Hamiltonian in Eq. \eqref{eq_gauge_invariant_H_deconf} is written as (up to a constant)
\begin{align}
    \mathcal{H}=&\kappa\sum_{\Box}\prod_{ij\in\Box}\rho^z_{ij}-\Gamma\sum_{i}\tau^z_{i}-\left(J+\frac{\Omega}{2}\right)\sum_{\langle ij\rangle} \tau^x_i\rho^z_{ij}\tau^x_j\nonumber\\
    &+\frac{\Omega}{4}\sum_i\sum_{\alpha={\bf \hat{x}},{\bf \hat{y}}}\left[\tau^x_i\rho^z_{i,i+\alpha}\rho_{i+\alpha,i+2\alpha}^z\tau^x_{i+2\alpha}\right].
    \label{eq_igtham}
\end{align}

Therefore in the limit of $\kappa\rightarrow\infty$, we have the $\pi$-flux constraint
\begin{align}
    \prod_{ij\in\Box}\rho^z_{ij}=-1~~~~~~\forall~~\Box
    \label{eq_frustration}
\end{align}

In terms of the IGT, the Ising electric charges, created by $\tau^x_i$ and obeying Gauss's law \eqref{eq_gaussising3}, hop on the sites of the dual square lattice (Fig. \ref{fig_dimer_model}) in a background $\pi$-flux through each plaquette. The amplitude for nearest neighbour (NN) and third nearest neighbour (3NN) hopping are $-(J+\Omega/2)$ and $\Omega/4$, respectively. Also note that $\rho^z_{ij}$ commutes with the Hamiltonian, such that the gauge fields do not have independent dynamics.

\subsection{\texorpdfstring{$\Omega=0$}{} line}

\begin{figure}
\includegraphics[scale=0.55]{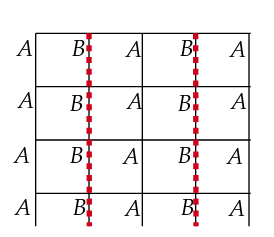}
 \caption{Schematic diagram for the choice of gauge to implement the $\pi$-flux condition  \eqref{eq_gaussising3} for the IGT. The dashed (solid) red (black) link corresponds to $\rho^z_{ij}=-1(+1)$ in Eq. \eqref{eq_igtham}. The two-site unit-cell consists of one site each from the two sublattices marked as $A$ and $B$.}
    \label{fig_2_site_unit_cell}
\end{figure}

For $\kappa\to\infty$ and $\Omega=0$, Eq. \eqref{eq_igtham} reduces to the NN fully frustrated transverse-field Ising model on the square lattice
\begin{align}
    \mathcal{H}_{\rm eff}=&-\Gamma\sum_{i}\tau^z_{i}-J\sum_{\langle ij\rangle} \tau^x_i\rho^z_{ij}\tau^x_j
    \label{eq_tfim}
\end{align}
where the frustration originates from $\pi$-flux condition in Eq. \eqref{eq_frustration}.  The phase diagram was obtained in Ref. \onlinecite{PhysRevB.30.1362}.

We choose a particular gauge (Fig. \ref{fig_2_site_unit_cell}) to implement the $\pi$-flux. For $J=0$, the ground state is an electric charge vacuum which is a topological liquid corresponding to the $\pi$-toric code. Note that the above gauge, on a 2-tori, chooses one of the topologically degenerate ground states of the toric code depending on the size of the system and boundary conditions. 

A finite $J$ leads to hopping of the electric charges while the magnetic charges are still gapped. In the above gauge, the soft modes of the dispersing electric charges can be obtained by diagonalising the hopping term within a soft spin analysis. Choosing a two-site unit cell as shown in the Fig. \ref{fig_2_site_unit_cell}, the soft mode quadratic Hamiltonian is given by 
\begin{align}
H_{\tau} =& \sum_{{\bf k}}\tau^x({\bf k})^T~M({\bf k})~\tau^x(-{\bf k})
\label{eq_softmodehamgauss}\\
\text{where}\\ &~~~
\tau^x({\bf k}) =  \begin{bmatrix}
      \tau^x_A({\bf k}) &  \tau^x_B({\bf k})
    \end{bmatrix}^T ,\\ \nonumber \\
    M({\bf k})&=  J~\begin{bmatrix}
        -2 \cos k_y & -(1+e^{2ik_x}) \\ 
        -(1+e^{-2ik_x}) & 2 \cos k_y 
    \end{bmatrix} 
\end{align}
Here, $k_x \in [-\pi/2,\pi/2]$ and $k_y \in [-\pi,\pi]$ belong to the magnetic Brillouin zone. The dispersion of the two eigenmodes of the above Hamiltonian is
\begin{align}
    E_{\tau}({\bf k})_{\pm} =\pm 2 J  \sqrt{\cos^2(k_x)+\cos^2(k_y)}
    \label{eq_softdispersion}
\end{align}
which has minima at 
$(k_x,k_y)$ = (0,0) and (0,$\pi$). The corresponding (un-normalised) eigenvectors at these minima are
\begin{align}
    \lambda_{\{0,0\}} = \begin{bmatrix}
        1+\sqrt{2} \\ 1 
    \end{bmatrix}  ~~,~~ \lambda_{\{0,\pi\}} = \begin{bmatrix}
        1 \\ 1 +\sqrt{2}
    \end{bmatrix}
\end{align}

Therefore, the explicit  soft-mode expansion of the electric charges is given by
\begin{align}
\left[\begin{array}{c}
    \tau^x_A(x,y)\\
    \tau^x_B(x,y)\\
\end{array}\right]= \phi_1 (x,y)~\lambda_{\{0,0\}}+ \phi_2 (x,y)\,e^{i\pi y}~\lambda_{\{0,\pi\}}
\label{eq:soft_mode_expansion}
\end{align}
where $\phi_1$ and $\phi_2$ are the corresponding amplitudes. Accounting for the projective representation, the symmetry transformation of the two soft modes under various lattice symmetries is given in Table \ref{tab_symproj}. 

\begin{table}
\begin{tabular}{|c|c|c|c|c|c|c|} \hline 
 $\phi$ & $T_x$ & $T_y$ & $R_{\pi}$ & ${\rm Ref}_x$ & $C_4$ & TR  \\ \hline  $\phi_1$ & $\phi_2$& $\phi_1$ & $\phi_1$ & $\phi_1$ & $\frac{\phi_1+\phi_2}{\sqrt{2}}$ & $\phi_1$ \\  
\hline $\phi_2$ & $\phi_1$& -$\phi_2$ & $\phi_2$ & $\phi_2$ & $\frac{\phi_1-\phi_2}{\sqrt{2}}$& $\phi_2$ \\
\hline\hline
$\Phi=\phi_1+i\phi_2$ & $i \Phi^*$ & $\Phi^*$ & $\Phi$ & $\Phi$ & $e^{i\pi/4}\Phi^*$ & $\Phi^*$\\ \hline
    \end{tabular} 
    \caption{The transformation of the two soft modes under square lattice symmetries: $T_x (T_y)$ is unit lattice translation along the $x (y)$ direction, $R_{\pi}$ represents a rotation by $\pi$ about the site, ${\rm Ref}_x$ corresponds to a reflection about the $y$ axis, $C_4$ denotes the $4$-fold rotation about a lattice site and TR is time reversal.}
    \label{tab_symproj}
\end{table}

\subsection{\texorpdfstring{$\Omega\neq 0$}{}}
\label{appen_sfotmode_omega}

At finite $\Omega$, the soft mode analysis for the quadratic Hamiltonian is given by Eq. \eqref{eq_softmodehamgauss} where now 
\begin{align}
    M(\Omega,{\bf k})=&  \left(J+\Omega/2\right)~\begin{bmatrix}
        -2 \cos k_y & -(1+e^{2ik_x}) \\ 
        -(1+e^{-2ik_x}) & 2 \cos k_y 
        \end{bmatrix}\nonumber\\
        &+\frac{\Omega (\cos(2k_x)+\cos(2k_y))}{2}\begin{bmatrix}
            1 & 0\\
            0 & 1\\
        \end{bmatrix}
\end{align}
The spectrum is given by (instead of Eq. \eqref{eq_softdispersion})
\begin{align}
    E_\tau(\Omega, {\bf k})_\pm=&
\frac{\Omega}{2}\left[\cos(2k_x)+\cos(2k_y)\right]\nonumber\\
&\pm
2\left(J+\frac{\Omega}{2}\right)
\sqrt{\cos^2 k_x+\cos^2 k_y}.
\end{align}
The two minima at $(0,0)$ and $(0,\pi)$ survive for
\begin{align}
    \frac{\Omega}{J}<\frac{2}{2\sqrt{2}-1}\approx 1.09
\end{align}
Above this, there is a continuous set of minima at $(Q_x, Q_y)$ satisfying 
\begin{align}
    \cos^2 Q_x + \cos^2 Q_y=\left(\frac{J+\Omega/2}{\Omega}\right)^2
\end{align}
such that just above the critical value, {\it i.e.}, $\Omega/J=\frac{2}{2\sqrt{2}-1}+\delta$, the set of minima consists of a circle of radius $\sim \sqrt{\delta}$ around $(0,0)$ or $(0,\pi)$ for $\delta\ll 1$. Hence, the momentum and, therefore, the tilt changes continuously within the quadratic theory.

\section{Derivation of the field theory from the soft modes}
\label{appen_qft}

The soft modes (Eq. \eqref{eq_orderp}) obtained from the analysis of the IGT (Appendix \ref{appen_z2gt}) leads to the low energy Lattice action (on the three dimensional cubic lattice in Euclidean space-time) for the soft modes allowed by the PSG (Table \ref{tab_symproj})
\begin{align}
    \mathcal{S}_{\mathbb{Z}_2}=&-t\sum_{\langle ij\rangle}\rho^z_{ij}\cos\left(\theta_i-\theta_j\right)-w\sum_i\cos(8\theta_i)\nonumber\\
    &+\frac{i\pi}{4}(1-\rho_{ij}^z)\left(1-\prod_{ij\in\Box}\mu^z_{IJ}\right)
\end{align}
where we have used the lattice version of Eq. \eqref{eq_softphase}. The first term denotes the hopping of the soft (electric mode) in the background Ising gauge field $\rho_{ij}^z$ while the second term is the lowest symmetry allowed term~\cite{PhysRevB.30.1362,PhysRevB.102.235124,PhysRevB.62.7850,PhysRevB.84.104430} that reduces the $O(2)$ symmetry of the soft mode action. Finally, the third term corresponds to the Ising Chern-Simons term~\cite{PhysRevB.62.7850,PhysRevB.84.104430} that captures the mutual semionic statistics between the electric charges and (gapped) magnetic charges with the latter being minimally coupled to the dual Ising gauge field $\mu^z_{IJ}$ on the links of the dual lattice ({\it i.e.}, the lattice on which the dimers sit). 

The partition function is given by
\begin{align}
    \mathcal{Z}=\sum_{\{\rho^z\}}\sum_{\{\mu^z\}}\int_{-\pi}^{\pi} \mathcal{D}\theta~e^{-\mathcal{S}_{\mathbb{Z}_2}}
\end{align}

Following Refs. \onlinecite{PhysRevB.62.7850,PhysRevB.102.235124,PhysRevB.84.104430} we use Villain transformation and Poisson re-summation formula to convert the partition function to
\begin{align}
    \mathcal{Z}=\sum_{\{\rho^z\}}\sum_{\{\mu^z\}}\sum_{\{L\}}\int_{-\pi}^{\pi} \mathcal{D}\theta~e^{-\mathcal{S}^{(1)}_{\mathbb{Z}_2}}
\end{align}
with the low-energy action
\begin{align}
    \mathcal{S}^{(1)}_{\mathbb{Z}_2}=&\sum_{\langle ij\rangle}\frac{L_{ij}^2}{2t}+i\sum_{\langle ij\rangle}L_{ij}\left(\theta_i-\theta_j+\frac{\pi}{2}(1-\rho^z_{ij})\right)\nonumber\\
    &-w\sum_i\cos(8\theta_i)+\frac{i\pi}{4}(1-\rho_{ij}^z)\left(1-\prod_{ij\in\Box}\mu^z_{IJ}\right)
    \label{eq_actionstep}
\end{align}
where $L_{ij}\in \mathbb{Z}$ is an integer-valued field defined on the links of the lattice.

Temporarily setting $w=0$, we can integrate out $\theta_i$ to constrain 
\begin{align}
    \sum_{j\in i}L_{ij}=0~~~~~\forall~i
\end{align}
which can be solved by introducing an integer dual gauge field, $b_{IJ}$, on the links of the dual lattice via the lattice curl
\begin{align}
    L_{ij}=\nabla\times b_{IJ}
\end{align}
such that the partition function is given by 
\begin{align}
    \mathcal{Z}=\sum_{\{\rho^z\}}\sum_{\{\mu^z\}}\sum_{\{b\}}\int_{-\pi}^{\pi} \mathcal{D}\theta~e^{-\mathcal{S}^{(2)}_{\mathbb{Z}_2}}
\end{align}
where the action (after rearranging) is
\begin{align}
    \mathcal{S}^{(2)}_{\mathbb{Z}_2}=&\sum_{\Box}\frac{(\nabla\times b_{IJ})^2}{2t}\nonumber\\
    &+i\frac{\pi}{2}\sum_{\langle ij\rangle}(1-\rho^z_{ij})\left[(\nabla\times b_{IJ})+\left[1-\prod_{ij\in\Box}\mu^z_{IJ}\right]\right].
\end{align}
Now, $\rho^z_{ij}$ can be integrated individually on each link of the direct lattice separately to obtain the constraint 
\begin{align}
    (-1)^{(\nabla\times b_{IJ})}=\prod_{ij\in\Box}\mu^z_{IJ}
\end{align}
Continuing to follow methods used in Refs. \onlinecite{PhysRevB.62.7850,PhysRevB.84.104430}, we divide the integer gauge field into two parts
\begin{align}
    b_{IJ}=2B_{IJ}+\beta_{IJ}
\end{align}
where $B\in\mathbb{Z}$ keeps track of the even part and $\beta_{IJ}=0,1$ is the odd part. Therefore, the above constraint becomes 
\begin{align}
    \mu_{IJ}^z=1-2\beta_{IJ}
\end{align}
Implementing the Integer constraint on the gauge field $b_{IJ}$ softly via the potential
\begin{align}
    -\mathcal{J}\sum_{\langle IJ\rangle}\cos(2\pi B_{IJ})=-\mathcal{J}\sum_{\langle IJ\rangle}\mu^z_{IJ}~\cos(\pi b_{IJ})
\end{align}
(with $\mathcal{J}>0$) we obtain
\begin{align}
    \mathcal{Z}=\sum_{\{\mu^z_{IJ}\}}\int \mathcal{D}b~e^{-\mathcal{S}^{(3)}_{\mathbb{Z}_2}}
\end{align}
where 
\begin{align}
    \mathcal{S}^{(3)}_{\mathbb{Z}_2}=\frac{1}{2t\pi^2}\sum_{\Box}\left(\nabla\times b_{IJ}\right)^2-\mathcal{J}\sum_{\langle IJ\rangle}\mu^z_{IJ}\cos(b_{IJ}).
\end{align}
We have shifted $b_{IJ}\rightarrow b_{IJ}/\pi$. We now integrate out $\mu^z_{IJ}$ per bond to obtain
\begin{align}
    \mathcal{Z}=\int \mathcal{D}b~e^{-\mathcal{S}^{(4)}_{\mathbb{Z}_2}}
\end{align}
where
\begin{align}
    \mathcal{S}^{(4)}_{\mathbb{Z}_2}
    &=\sum_{\Box}\frac{\left(\nabla\times b_{IJ}\right)^2}{2t\pi^2}-\sum_{\langle IJ\rangle}\mathcal{J}^2\cos(2b_{IJ})+\cdots
\end{align}
 We can now extract a charge field, $\chi_I$, by choosing a gauge $\nabla\cdot b=0$
such that we get
\begin{align}
    \mathcal{Z}=\int \mathcal{D}b~e^{-\mathcal{S}^{(5)}_{\mathbb{Z}_2}}
\end{align}
where
\begin{align}
    \mathcal{S}^{(4)}_{\mathbb{Z}_2}&=\sum_{\Box}\frac{\left(\nabla\times b_{IJ}\right)^2}{2t\pi^2}-\mathcal{J}^2\sum_{\langle IJ\rangle}\cos(\chi_I-\chi_J+2b_{IJ})\nonumber\\
    &=\sum_{\Box}\frac{\left(\nabla\times b_{IJ}\right)^2}{2t\pi^2}-\mathcal{J}^2\sum_{\langle IJ\rangle}\left[\Psi^*_I e^{i2b_{IJ}}\Psi_J+{\rm c.c.}\right]
\end{align}
with 
\begin{align}
    \Psi_I=e^{-i\chi_I}
\end{align}
being the charge-$2$ boson on the sites of the lattice dual to the soft electric modes.

The continuum limit can now be taken to obtain
\begin{align}
    \mathcal{Z}=\int \mathcal{D}^2\Psi~\mathcal{D}b~e^{-\mathcal{S}^{(6)}_{\mathbb{Z}_2}}
\end{align}
where
\begin{align}
    \mathcal{S}^{(6)}_{\mathbb{Z}_2}=&\int d^2{\bf x}d\tau\left[\frac{1}{2}\left|\left(\partial_\mu-i2b\right)\Psi\right|^2-r|\Psi|^2+u|\Psi|^2\right]\nonumber\\
    &+\int d^2{\bf x}d\tau\left[\frac{1}{2}(\partial_0{\bf b}-\nabla b_0)^2+\frac{1}{2}(\nabla\times {\bf b})^2\right].
    \label{eq_abelianhiggs2e}
\end{align}
This is the Abelian Higgs model with charge-$2$ bosons obtained in the description of the $\mathbb{Z}_2$ liquids in dimer models on non-bipartite latices~\cite{sachdev1999translational,PhysRevB.42.4568}.

It is convenient to dualise the above action once more with respect to the bosons, $\Psi$, to re-write it in terms of the soft electric modes (Eq. \eqref{eq_orderp} in the main text) as can be verified~\cite{PhysRevB.102.235124,PhysRevB.79.064405} by applying bosonic particle-vortex duality~\cite{peskin1978mandelstam,dasgupta1981phase} to $\Psi$, whence we get
\begin{align}
    \tilde{\mathcal{S}}=&\int d^2{\bf x}d\tau~\left[\frac{1}{2}\left|\left(\partial_\mu-ia_\mu\right)\Phi\right|^2+\frac{r}{2}|\Phi|^2+u|\Phi|^4\right]\nonumber\\
    &+\int d^2{\bf x}d\tau~\left[i\frac{1}{\pi}\epsilon^{\mu\nu\lambda}a_\mu\partial_\nu b_\lambda\right]\nonumber\\
    &+\int d^2{\bf x}d\tau \left[\frac{1}{2}(\partial_0{\bf b}-\nabla b_0)^2+\frac{1}{2}(\nabla\times {\bf b})^2\right]\nonumber\\
    &+\int d^2{\bf x}d\tau \left[\frac{1}{2}(\partial_0{\bf a}-\nabla a_0)^2+\frac{1}{2}(\nabla\times {\bf a})^2\right]
\end{align}
where $\Phi$ us dual to $\Psi$ and hence reflects the electric charge modes. This is exactly equal to the action given by Eq. \eqref{eq_action} in the main text, except for the last term, which is the same as turning on the $w\neq 0$ term in Eq. \eqref{eq_actionstep}. This completes our derivation of the continuum action.

Finally, we note that the symmetry transformation of the gauge fields under the microscopic symmetries is fully fixed by their minimal coupling to the matter field, $\Phi$ and its dual. In particular, the transformation of $a$ is given by (see Table \ref{tab_symproj} for notations) is given by
\begin{align}
    T_x, T_y~:&~a_\mu\rightarrow -a_\mu~~~\mu=0,1,2\nonumber\\
    R_\pi~:&~a_0\rightarrow a_0, a_1\rightarrow -a_1, a_2\rightarrow -a_2\nonumber\\
    {\rm Ref}_x~:&~a_0\rightarrow a_0, a_1\rightarrow -a_1, a_2\rightarrow a_2\\
    C_4~:&~a_0\rightarrow -a_0, a_1\rightarrow -a_2, a_2\rightarrow a_1\nonumber\\
    {\rm TR} ~:&~a_0\rightarrow-a_0, a_1\rightarrow a_1, a_2\rightarrow a_2\nonumber
\end{align}

\subsection{The equal time correlator for the \texorpdfstring{$z=2$}{} Gaussian theory in Eq. \ref{eq_gaussianlifshitz}}
\label{appen_correlator}

The Gaussian theory in Eq. \eqref{eq_gaussianlifshitz} in Fourier space is given by
\begin{align}
    \tilde{\mathcal{S}}_g=&\frac{1}{2}\int \frac{d^2{\bf k}}{4\pi^2}\frac{d\omega}{2\pi}\left[\left(\omega^2+\eta_2{\bf k}^4+\lambda_2k_1^2k_2^2\right)\right]\theta_{{\bf k},\omega}\theta_{-{\bf k},-\omega}
    \label{eq_gaussianlifshitz2}
\end{align}

Therefore, the equal time correlator including the anisotropy can be calculated as an extension of the same done in Ref. \onlinecite{PhysRevB.69.224416} as follows.

\begin{align}
    \langle\theta_{\bf x}\theta_0\rangle&=\int \frac{d^2{\bf k}}{4\pi^2}\frac{d\omega}{2\pi} e^{i\bf k\cdot x}\langle \theta_{{\bf k},\omega}\theta_{-{\bf k},-\omega}\rangle\nonumber\\
    &=\int \frac{d^2{\bf k}}{4\pi^2}\frac{d\omega}{2\pi} e^{i\bf k\cdot x}~\frac{1}{\omega^2+\eta_2{\bf k}^4+\lambda_2k_1^2k_2^2}\nonumber\\
    &=\frac{1}{2\pi^2\sqrt{\eta_2}\sqrt{1+\epsilon/4}}\mathcal{K}\left(\frac{\epsilon}{4+\epsilon}\right)\nonumber\\
    &~~~\times\sum_{m=-\infty}^\infty \left[\frac{1-\sqrt{1+\epsilon/4}}{1+\sqrt{1+\epsilon/4}}\right]^{|m|}\int_0^\infty \frac{d{k}}{k}\mathcal{J}_{4m}(kx)
\end{align}
where 
\begin{align}
    \mathcal{K}(a)=\int_0^{\pi/2}\frac{d\theta}{\sqrt{1-a\sin^2\theta}}
\end{align}
is the complete elliptical integral of the first kind and $\mathcal{J}_m(a)$ is the Bessel function. Using
\begin{align}
    \int_0^\infty \frac{d{k}}{k}\mathcal{J}_{4m}(kx)=\left\{\begin{array}{ll}
    \frac{1}{4m} & m\neq 0\\
    -\ln x + \cdots & m=0\\
    \end{array}\right.
\end{align}
we get, to the leading order Eq. \eqref{eq_monopolecorrelator} in the main text.
\bibliography{ref}

\end{document}